\documentclass[12pt]{iopart}

\usepackage{graphicx}
\usepackage{dcolumn}
\usepackage{bm}
\usepackage{tabularx}
\usepackage{booktabs}

\usepackage{iopams}
\begin{document}
\bibliographystyle{unsrt}
\title{Natural and artificial atoms for quantum computation}

\author{Iulia Buluta$^1$, Sahel Ashhab$^{1,2}$, Franco Nori$^{1,2}$}

\address{
$^1$Advanced Science Institute, RIKEN, Wako-shi, Saitama, 351-0198, Japan}
\address{
$^2$Department of Physics, The University of Michigan, Ann Arbor, Michigan 48109-1040, USA}


\begin{abstract}
Remarkable progress towards realizing quantum computation has been achieved using natural and artificial atoms as qubits. This article presents a brief overview of the current status of different types of qubits. On the one hand, natural atoms (such as neutral atoms and ions) have long coherence times, and could be stored in large arrays, providing ideal ``quantum memories''. On the other hand, artificial atoms (such as superconducting circuits or semiconductor quantum dots) have the advantage of custom-designed features and could be used as ``quantum processing units''. Natural and artificial atoms can be coupled with each other and can also be interfaced with photons for long-distance communications. Hybrid devices made of natural/artificial atoms and photons may provide the next-generation design for quantum computers.
\end{abstract}

\pacs{03.67.Lx}
\maketitle

\section{Introduction}
The experimental realization of Quantum Computation (QC) has been a challenge for more than a decade. While a fully operational quantum computer that could factorize thousand-digit numbers is still a distant goal, with the new technologies for the coherent manipulation of atoms, photons, and electrons, nowadays applications like quantum cryptography and quantum communication are already commercially available. Since potential QC implementations come in many shapes and sizes, it is difficult to quantify the overall progress in the field of QC. In order to assess the current state of the art in QC, a comparison between the various approaches is needed. However, because these approaches are very different (in terms of the underlying physical processes, experimental techniques, and how well the physical system is understood), we should be careful not to compare apples with oranges.  We would rather like to compare apples with apples, or in our case, atoms with atoms. Therefore, in this paper we consider natural and artificial atoms for implementing QC.
\par
Among the most successful and rapidly developing ways of realizing QC are those using {\it natural atoms}  (such as neutral atoms \cite{Bloch08} or ions \cite{Bla08}) and {\it artificial atoms}  (such as superconducting circuits \cite{CW08,You05} or spins in solids \cite{HA08}). Contrasting natural and artificial atoms would help highlighting their strengths. For the sake of comprehensiveness other QC approaches (i.e., with nuclear spins in molecules \cite{VC05,Bau07} or in phosphorus impurities in silicon \cite{Kan98,Mor08}, photons \cite{Gis07,Kok07}, and so on)  are also be briefly covered here.  A complementary overview on qubits can be found in \cite{Lad10}. Although there are many exciting theoretical proposals, we will focus more on what has already been experimentally demonstrated and less on what could eventually be achieved in each system. We should stress from the beginning that our purpose is not to show that a certain system is better than others, but to review the current experimental state of the art in QC.  One should keep in mind that  some approaches are more recent than others, some benefit from technologies that have been developed before, while others had to develop their own new technologies on the way, and, most importantly, each approach has to deal with specific issues whose difficulty cannot be compared.
\par
By considering natural and artificial atoms and their potential for implementing QC, we hope to gain a broader perspective of the current status of QC. Moreover, this approach may also provide a glimpse into  the future of QC.  However, we would rather not attempt to make any prediction regarding what system would be best for realizing a practical quantum computer. Ten or twenty years from now such speculation might sound as amusing as the prediction made by  \textit{Popular mechanics} in 1949: ``In the future, computers may weigh no more than 1.5 tonnes.''
\par
After summarizing the characteristics of each system we discuss the strengths and weaknesses of natural and artificial atoms. Next, we take a look at hybrid systems and photon interfaces, and, finally, consider future prospects. The main issues discussed throughout the paper are collected in the six tables, which can be found at the end of the paper. For the reader interested in the details for a particular system, the Appendix provides extended tables. The list of references at the end tries to cover some of the recent experimental progress in the coherent control of natural and artificial atoms.

\section{Neutral atoms}

When looking for a physical system to realize qubits (which are controllable two-level systems), perhaps the most obvious candidate is neutral atoms \cite{Man03,Sch04,Tre04,Jak05,Mic06,Mir06,Yav06,And07,Beu07,Hay07,Lew07,Nel07,Tro08,Gae09,Saff09,Urb09,Wur09,Deu10,Fuh10,Gib10,Ise10,Olm10,She10,Wei10,Wil10,Zha10}. Atoms have many energy levels that have been studied extensively over the past century, and some of these energy levels are extremely stable. Indeed, with accuracies better
than one part in $10^{-15}$, atomic clocks provide the best available time and frequency standards. The qubits encoded in the atomic energy levels can be initialized by optical pumping and laser cooling, manipulated with electromagnetic radiation, and then measured via laser-induced fluorescence. In short, atoms provide clean, well-defined qubits (see also Box 2 (a,b) and Table \ref{at}).
\par
Neutral atoms make attractive qubit candidates also because of their weak interaction with the environment, leading to long coherence times \cite{Tre04,Sch04,Yav06,Deu10}.  They can be cooled down to nK temperatures and trapped in very large numbers (millions) in microscopic arrays created by laser beams (called optical lattices). The trapping and manipulation of atoms can be done with high precision \cite{Sch04,Mir06,Yav06,Beu07}.  Until recently, the individual manipulation and measurement of neutral atoms in optical lattices was not possible, but the experiments in \cite{Nel07,Wur09,She10,Gib10,Fuh10} show very promising perspectives for individual addressing and readout.
\par
While one-qubit gates can be implemented with very high fidelity \cite{Olm10}, realizing two-qubit gates or many-qubit entangled states is challenging because the atoms interact very weakly with each other. This problem can be overcome in several ways. For instance, the atoms can first be brought into a superposition of two internal spin states. Then, as the spin-dependent lattice is moved, the atoms go to the left and to the right simultaneously colliding with their neighbors. In this way, in a single operation a highly entangled many-qubit state can be created  \cite{Man03}. Unfortunately, these collisional gates are very sensitive to decoherence and are also quite slow \cite{Bloch08}. Exchange interactions provide an alternative approach \cite{And07,Hay07,Tro08}. The effective spin-spin interaction between two atoms in a double-well potential was used to demonstrate a two-qubit SWAP gate \cite{And07}. Furthermore, with polar molecules \cite{Mic06} or Rydberg atoms \cite{Urb09,Saff09,Wei10} dipole-dipole interactions could be exploited for realizing two-qubit gates.  Very recently, a CNOT gate \cite{Ise10}, post-selective entanglement of two atoms \cite{Wil10} using Rydberg blockade interactions and on-demand entanglement \cite{Zha10}  have been demonstrated.
\par
The prospect of producing many-qubit entangled states together with the possibility of single-site addressing and measurement make neutral atoms promising for the quantum simulation of condensed-matter physics \cite{Jak05,Lew07} as well as measurement-based QC \cite{Kay06}.

\section{Ions}

While neutral atoms interact weakly among themselves, ions, being charged, interact rather strongly via Coulomb repulsion. This facilitates the implementation of two-qubits gates without compromising the long coherence times \cite{Tur98,Cir00,Sac00,Sch03,Gul03,Lei03,Bar04,Chi04,Rie04,Chi05,Hae05,Haf05,Lei05,Rie06,Sti06,Moe07,Ben08,Kie08,Han09,Myr08,Mon09,Moz09,Bur10,Cam10,Mon10}. Also thanks to their charge, the motion and position of the ions can be well controlled. Ions can be trapped by electrical (or magnetic) fields, laser-cooled and manipulated with high precision \cite{Bla08}. Quantum information can be encoded either in the internal (hyperfine or Zeeman sublevels, or the ground and excited states of an optical transition), or in the motional states (the collective motion of the ions). While the internal states exhibit very long coherence times (hyperfine transitions $>20$ s \cite{Hae05} and optical transitions $>1$ s)  the motional states have typical lifetimes of $<100$ ms. As in the case of neutral atoms, the initialization of the qubits can be done by optical pumping and laser cooling, and they can be measured with very high accuracy \cite{Myr08,Bur10} via laser-induced fluorescence.  Scaling the current experiments to large numbers of ions is theoretically possible, but technically challenging. The proposed approaches to scalability include ion shuttling, two-dimensional ion arrays, photon interconnections, long equally-spaced strings, and two-dimensional Coulomb crystals (see \cite{Kie08} and Box 2 (c,d) and Table \ref{ions}).
\par
Using the collective motion of the ions as data bus, high-fidelity one-, two- \cite{Rie06,Ben08}, and even three-qubit \cite{Mon09} gates have been experimentally demonstrated. Entangled  (Greenberger-Horne-Zeilinger (GHZ) and W) states of up to 14 qubits have been realized \cite{Haf05,Lei05,Mon10}.
Two-qubit gates can also be  implemented using bichromatic excitation fields that produce coherent two-qubit transitions \cite{Sac00,Ben08} or by the state-selective displacement of the ions with an optical ``pushing'' force \cite{Cir00}. In the latter, the displacement changes the strength of the Coulomb repulsion, leading to an additional phase, so realizing a controlled-phase gate.  Recently, a trapped ion quantum processor implementing arbitrary unitary transformations on two qubits has been realized \cite{Han09}.
\par
Besides the generation of GHZ and W entangled states, quantum algorithms \cite{Gul03,Chi05}, quantum teleportation \cite{Bar04,Rie04}, entanglement of distant qubits \cite{Moe07},  quantum error correction \cite{Chi04} and decoherence free qubits \cite{Moz09} have also been demonstrated with trapped ion qubits.

\section{Superconducting circuits}

Superconducting circuits \cite{Nak99,Nak02,Mar02,Vio02,Chi03,Ber03,Yam03,Wal04,McD05,Him06,Gra06,Ste06,Yos06,Har07,Lup07,Maj07,Nis07,Pla07,Sil07,Pic08,Sch08,Sho08,Ans09,DiC09,Man09,Nee10,diC10,Sun10,Har10,OCo10,Pal10,Ste10,Mar11,Byl11,Martinis,Tyr11} are typically $\mu$m-scale circuits operated at mK temperatures. Although macroscopic, they can still exhibit quantum behavior, which can be harnessed for QC \cite{Mak01,You05,CW08,You11}. Superconducting circuits  are $RLC$ circuits that also include nonlinear elements, called Josephson junctions. Thanks to superconductivity, the resistance vanishes ($R = 0$), eliminating the most serious source of dissipation and noise.  Now, the $LC$ circuit is a harmonic oscillator. The problem with harmonic oscillators is that they have an infinite number of equally-spaced energy levels and therefore it is not possible to target only the lowest two energy levels. By introducing nonlinearity through the Josephson junction, the energy-level separation becomes nonuniform, and the lowest two levels can be used to encode the qubit \cite{You05,CW08} (see also Box 1). Quantum information can be encoded in different ways: in the number of superconducting electrons on a small island (charge qubit), in the direction of a current around a loop (flux qubit), or in oscillatory states of the circuit (phase qubit). 
These qubits can be controlled by microwaves, voltages, magnetic fields, and currents  as well as measured with high accuracy \cite{Pic08} using integrated on-chip instruments. The characteristics of the qubits can be designed and many qubits could be coupled in arrays. Therefore, superconducting qubits are flexible and promise the realization of QC on a  chip (see Box 2 (e,f) and Table \ref{sc}).
\par
Superconducting qubits have coherence times that can reach tens of $\mu$s (e.g., \cite{Byl11}), the coupling between qubits can be made strong and can be turned on and off electronically \cite{Him06,Nis07}. In addition to direct coupling strategies, superconducting circuits can be coupled via ``cavities'' \cite{Sil07,Maj07}, which are actually electrical resonators (and the ``photons'' are actually electron-density oscillations). This setup is promising  for the study of circuit cavity Quantum Electrodynamics (circuit QED) \cite{Wal04,Chi04,You05,CW08,Sho08}.
\par
With superconducting circuits one can now realize simple algorithms \cite{DiC09}, and generate entangled states of three qubits \cite{Nee10,diC10,Sun10} and arbitrary photon states in a resonator \cite{Hof09}. Other recent advances include the performance of quantum non-demolition measurements \cite{Lup07}, the realization of multi-level quantum systems \cite{Martinis,Nor09}, the violation of Bell's inequality \cite{Ans09,Pal10}, and the coupling of a mechanical resonator to a superconducting qubit \cite{OCo10}.

\section{Spins in solids}

Coherent control and measurement of single spins in solids \cite{Los98,Tyr03,Jel04,Jele04,Pet05,Chi06,Han06,Kop06,Steg06,Dut07,Han07,Mik07,Now07,Xu07,Ama08,Ber08,Chi08,Ger08,Neu08,Mor08,Bal09,Bar09,Fol09,Han09,Twa09,Bar10,Blu10,Lan10,Nad10,Blu11} is now possible, and this allows using electron spins in semiconductor quantum dots \cite{Han07}, or electron spins together with nuclear spins in nitrogen-vacancy (NV) color centers in diamond \cite{Dut07}  for QC purposes \cite{Los98,HA08} (see Box 2 (g,h) and Table \ref{spi} which attempts to cover, as much as possible in such a short space, several very different systems under the broad umbrella of spins).
\par
Quantum dots are nanoscale structures in which electrons are trapped in all three dimensions. They can be fabricated in several ways, for example, by growth or with electrode gates in a two-dimensional electron-gas. The material of choice is usually GaAs. On the other hand, NV centers are point defects in the diamond lattice, consisting of a nearest-neighbor pair made of a nitrogen atom, substituting a carbon atom, and a lattice vacancy.
Although in its early stages, quantum computing with electronic and nuclear spins in an array of phosphorus donor atoms embedded in a pure silicon lattice
(P:Si) has recently achieved very encouraging results \cite{Mor10,MCC10,Wit10,Sims10,Sim11}.
\par
Solid state qubits such as quantum dots are attractive because, like superconducting circuits, they could be designed to have certain characteristics and assembled in large arrays. Furthermore, they require temperatures of up to a few K (NV centers in diamond could operate even at room temperature). The manipulation and readout can be done both electrically \cite{Now07} and optically \cite{Xu07,Mik07,Ger08}.
\par
While Rabi oscillations have already been observed \cite{Kop06,Ber08}, two-qubit gates have only been demonstrated for NV centers in diamond \cite{Jele04} (although, a SWAP gate between logical states has been realized \cite{Pet05}). However, long coherence times \cite{Ama08,Chi08} have been measured for both quantum dots ($\sim \mu$s) \cite{Fol09,Blu10,Blu11,Bar09,Bar10} and NV centers ($>5$ ms) \cite{Neu08}. Moreover, for NV centers the entanglement between the electron and nuclear spins has also been shown \cite{Neu08}.
\par
Nowadays, Nuclear Magnetic Resonance (NMR) techniques are extensively used in the context of nuclear spins in semiconductors. NMR techniques have been used for the control of nuclear spins in molecules \cite{Bau07,Sut08,Pen05,VC05,Neg06}, which proved very successful for realizing QC with such nuclear spin qubits \cite{VC05,Bau07} (see also Table \ref{nmr}). A well-known example is the factorization of $N=15$ using Shor's algorithm \cite{Van01}. Nuclear spin qubits have long coherence times ($>1$ s) and high-fidelity quantum gates have been demonstrated \cite{VC05}. The coherent control of up to 12 qubits has also been realized \cite{Neg06}. However, this approach to QC proved difficult to scale up to tens or hundreds of qubits, so NMR techniques are now being applied for the control of nuclear spins in semiconductors.  One direction is solid-state NMR \cite{Sut08}, but  NMR is also merging with Electron Spin Resonance (ESR) methods, so it also becomes relevant for NV centers in diamond and for phosphorus in silicon QC.

\section{Comparing natural and artificial atoms}

The main characteristics of natural and artificial atoms are displayed in Tables 1 and 2. In Table 1:  $T_1$ (relaxation time) is the average time that the system takes for its excited state to decay to the ground state; $T_2$ (decoherence or dephasing time) represents the average time over which the qubit energy-level difference does not vary. We denote by $Q_1$ (quality factor) the number of one-qubit quantum gates that can be realized within the time $T_2$, and by $Q_2$ (quality factor) the number of two-qubit quantum gates that can be realized within the time $T_2$. For implementing QC we are mainly interested in the following aspects: {\it controllability}, {\it scalability} and {\it interface-ability}. The latter will also be discussed in the following section.
\par
\begin{figure}
\includegraphics[width=0.9\textwidth ]{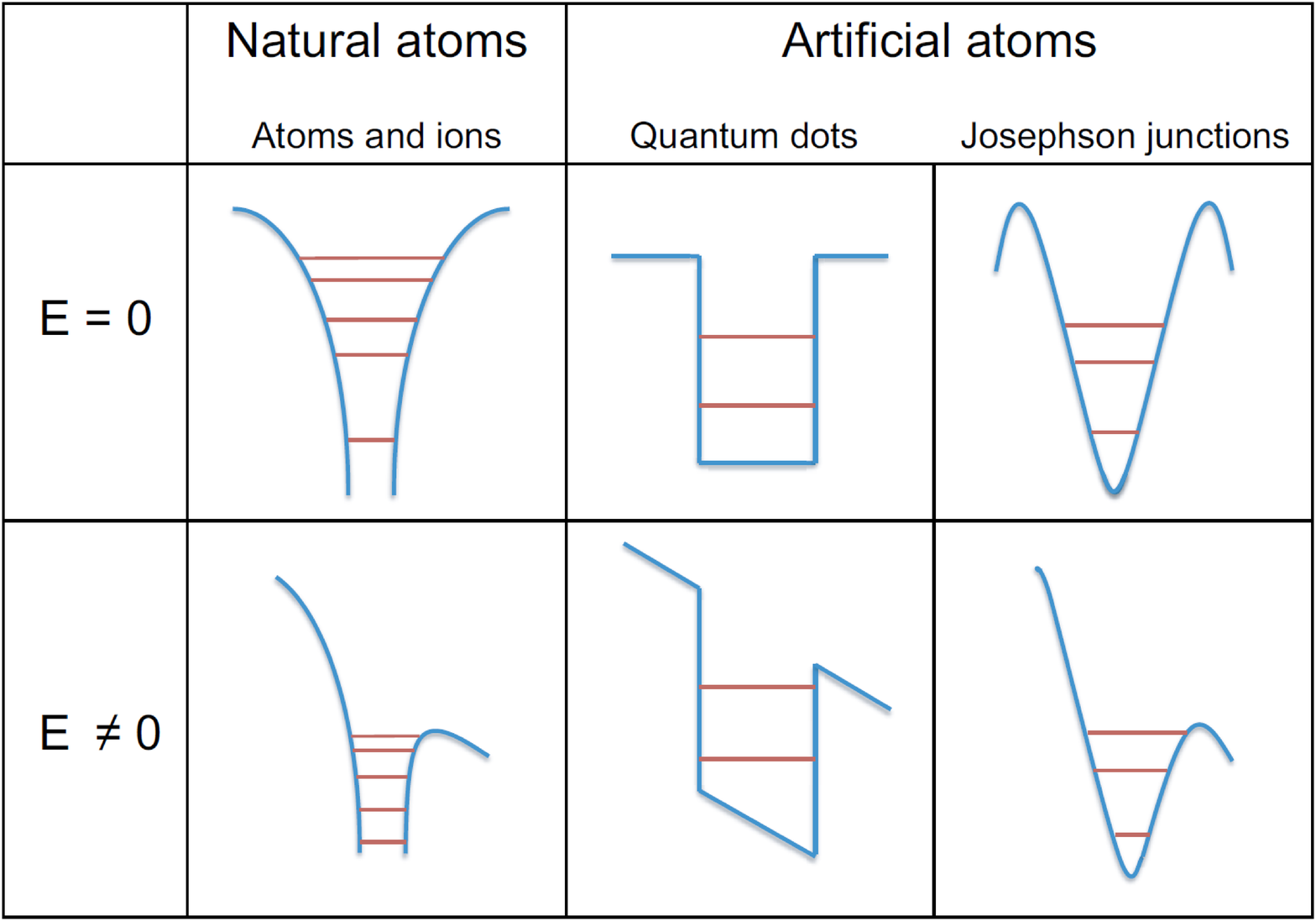}
\end{figure}

\framebox{\begin{minipage}[t]{\textwidth}%
\textbf{Box 1:  Natural and artificial atoms}
\\
Both natural and artificial atoms exhibit discrete energy levels, which are modified in the presence of external fields ($E \neq 0$). The applied external fields drive coherent quantum oscillations between the specific energy levels which can be used to encode the qubit states. Artificial atoms can be engineered to have certain transition frequencies while in natural atoms these are fixed.
\end{minipage}}
\\
\\
The qubit energy-level splittings are comparable for natural and artificial atoms -- microwave frequencies (for ions and superconducting circuits) and optical frequencies (for neutral atoms, ions and some semiconductor quantum dots). Box 1 displays schematically the potential energies and discrete energy levels for natural and artificial atoms in the absence ($E = 0$) and in the presence ($E \neq 0$) of an external field. While natural atoms are usually driven using optical or microwave radiation, artificial atoms like superconducting circuits can be driven by currents and voltages, magnetic fields, as well as microwave photons. Optically-driven artificial atoms, such as some semiconducting quantum dots, have also been demonstrated. Artificial atoms can be engineered to have a large dipole moment or particular transition frequencies. Depending on the intended application this tunability may prove quite useful.
\par
In natural atoms, motional states can also be exploited for encoding the qubits or as data bus. The motional frequency can be controlled, but the cooling of these modes is usually necessary if they are to be used for QC purposes. For artificial atoms, resonators can play a similar role to the motional modes. The frequency of these resonators can also be controlled, and they can be cooled much like atoms.  For instance, the temperature of superconducting circuits can be decreased using cooling techniques inspired from atomic physics, such as sideband or Sisyphus cooling \cite{Gra08,Nor08}. Natural atoms have many energy levels which can be used to encode information. Levels that are well-protected against decoherence (i.e., magnetic-field-independent hyperfine transitions \cite{Lan05}) could be used for memory qubits, while fast transitions could be used for implementing two-qubit gates. Furthermore, realizing qudits in natural atoms is straightforward.
\par
Unlike natural atoms of the same species, which are indistinguishable, no two artificial atoms will be perfectly alike. With the latest advances in microfabrication, artificial atoms can be made with increasing accuracy and uniformity.  However, this is an extra challenge. While natural atoms are readily available and one only needs to trap them by means of optical or electrical fields and then cool them down to low temperatures, artificial atoms have to be carefully designed and fabricated. Furthermore, atom and ion trapping technologies have been in use for quite a while, but for artificial atoms the techniques are more recent.
\par
Artificial atoms can be produced in large numbers and ``wired'' together on a chip. Therefore, extending current experiments to large numbers of artificial atoms should, in principle, not be a problem. Neutral atoms can be loaded by thousands or millions in optical lattices; however, individual addressing has not yet been fully demonstrated \cite{Wur09}.  Meanwhile, in the case of ions, although several proposals are available, scaling to large numbers is a challenge. Natural atoms are not wired so they can form almost any 2D or 3D configuration; however, for artificial atoms the wiring itself may impose some geometric limitations. Neutral atoms and trapped ions qubits can also be moved around easily. This flexibility may prove advantageous for certain applications.
\par
Both natural and artificial atoms can be coupled with photons via cavities QED \cite{You05,CW08,Sho08}, which could provide a means of realizing large scale QC and long distance quantum communication (see also \cite{Kim08}). The physics of cavity QED is the same regardless of the nature of the atom or cavity, but, for artificial atoms (e.g., circuit QED) the coupling strength is several orders of magnitude larger than for natural atoms \cite{You05,CW08,Sho08}. Several exciting experiments demonstrating the coupling between cavities and natural or artificial atoms have been performed (see, for instance, \cite{Col07,Her09,Eng07,Sil07,Maj07} and the review in \cite{You11}).
\par
As for the operating conditions, natural atoms can be coherently manipulated only in ultrahigh-vacuum at very low temperatures (nK-$\mu$K for neutral atoms and mK for ions). Artificial atoms are also operated at low temperatures (mK in the case of superconducting circuits or a few K for semiconductor quantum dots), but there are some candidates for room-temperature qubits, including very long coherence times for NV centers in diamond (note that their $T_1$  is temperature dependent).

\begin{figure}
  \includegraphics[width=\textwidth]{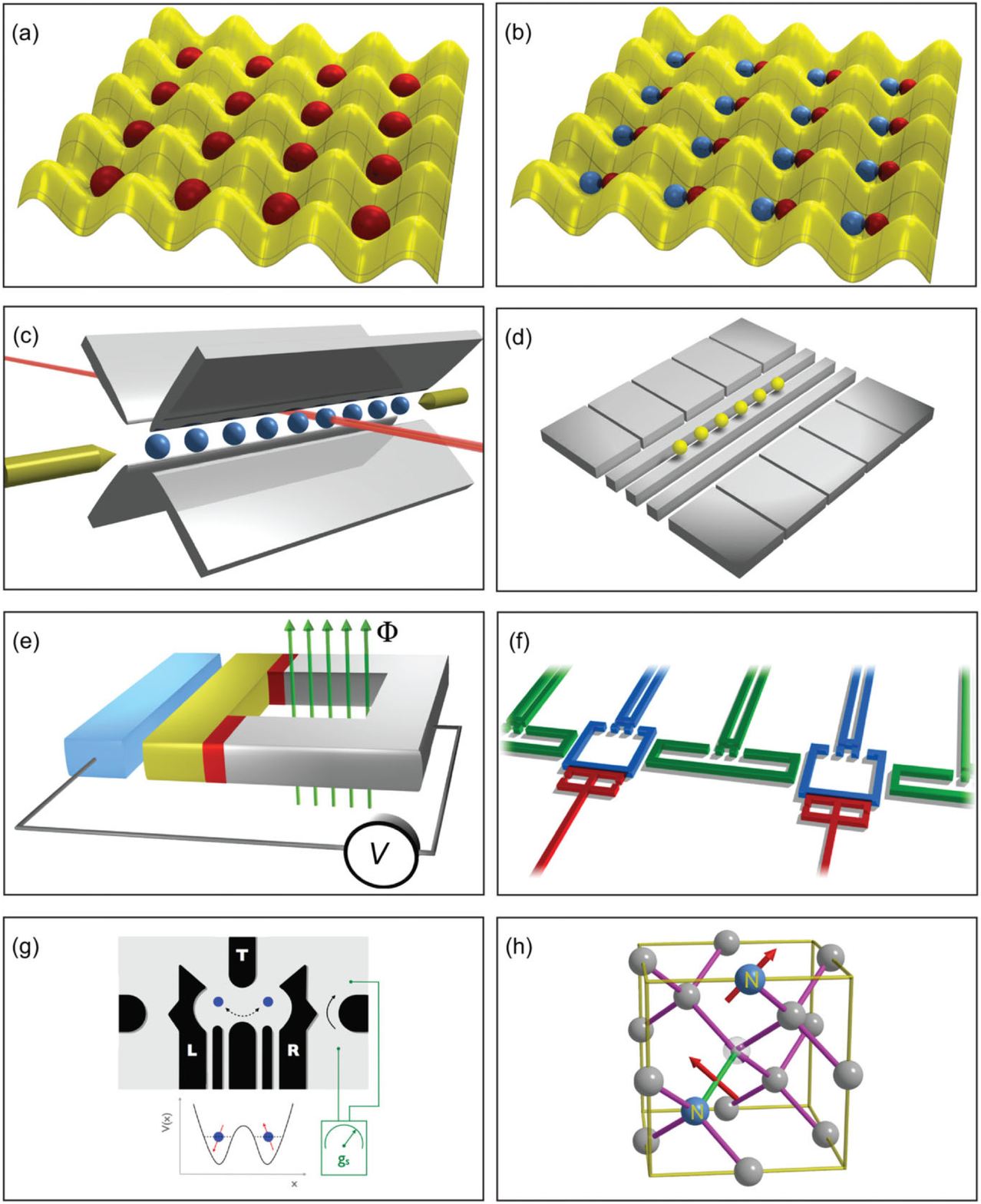}
\end{figure}

\framebox{\begin{minipage}[t]{\textwidth}%
\textbf{Box 2: Quantum bits}
\\
\\
Quantum bits can be constructed using a variety of different possible
building blocks, of various sizes and properties. As a result, each
technology has its unique advantages and challenges.
\\
\\
\textbf{(a,b)} Hundreds of thousands of neutral atoms can be trapped and cooled at
the minima of an optical lattice - the periodic potential created by
interfering counter-propagating laser beams. The long-lived internal energy
levels of neutral atoms are used to encode quantum information. Neutral atom
qubits can be manipulated with laser radiation and observed via their
laser-induced fluorescence. The typical separation between lattice sites is
$<1$ $\mu$m, which makes individual addressing challenging.
Neutral atoms interact weakly with the environment,  which protects them
from decoherence. There are several mechanisms for entangling neutral atoms:
through state-dependent displacement of the lattice, that results in a
highly entangled many-qubit state created in a single operation; through
exchange interactions; or via the interaction between two atoms in a
double-well potential. Neutral atoms in optical lattices are ideal systems
for quantum simulation.
(a) illustrates the idea of trapping neutral atoms in periodic optical potentials; one neutral atom - qubit is trapped at each lattice site;
(b) shows one possible mechanism for creating multi-particle entanglement starting with two atoms in different spin states, trapped in each lattice site.
\\
\\
\textbf{(c,d)} Ions trapped in electro-magnetic fields have been used to encode and
manipulate quantum information. The internal energy levels representing the qubit basis states are long-lived and can be easily excited with laser radiation. The
typical distance between trapped ions is 5 $\mu$m or more which
facilitates addressing and readout of individual ions. High-efficiency
readout is achieved by monitoring the laser-induced fluorescence. Ions in
the same potential have a common center-of-mass vibrational mode that can be
used as data bus to realize entangling operations. Many-particle
entanglement and high-fidelity two-qubit gates have already been
demonstrated in experiment. Panel (c) shows a linear trap, while (d) a
planar trap. These recently developed micrometer-scale ion traps (d) provide
flexibility in manipulating the positions of the ions in two and three
dimensions. Nowadays the main focus is on scaling these experiments to large
numbers of ions. This can be achieved by moving the ions in the trapping
potentials around in complex microstructures, trapping single ions at
specific locations in custom-designed lattice geometries created in arrays
of microtraps, or by entangling the ions with flying qubits (photons).
\\
\end{minipage}}

\framebox{\begin{minipage}[t]{\textwidth}
\textbf{Box 2:  Quantum bits (Continued)}
\\
\\
\textbf{(e,f)} Superconducting qubits are micrometer-sized electric circuits based on
Josephson junctions. A superconducting qubit (e) can be manipulated using
the applied electric voltage $V$ and magnetic flux $\Phi$. Similarly, the
qubit can be read out through the small electric or magnetic signal that it
produces. Additional circuit elements, called couplers, can be used to
provide tunable interactions between the qubits, as shown in (f), allowing
the creation of entanglement and the performance of two-qubit gates.
Decoherence times have improved from the nanosecond to the microsecond scale
over the past decade and are expected to improve further in the future.
\\
\\
\textbf{(g,h)} Spins in solids arise in a number of distinct realizations. The
collective spin state of two electrons trapped in a sub-micrometer-scale
semiconductor-based double quantum dot structure can be used as a qubit, as
shown in (g). In the traditional approach, magnetic fields are used to
manipulate the qubit, but recent techniques using electric fields and
exploiting the exchange and spin-orbit interactions have been developed as
well. The qubit is readout by monitoring its response to an applied electric
signal. Nitrogen-vacancy (NV) centers in diamond, shown in (h), also provide
alternative spin qubits. The spin of one electron in the NV chemical bond
can be manipulated and read out using magnetic fields and optical-frequency
electromagnetic fields. These qubits have long coherence times, on the
millisecond timescale. It would be highly desirable to controllably place
multiple qubits in an ordered arrangement in the diamond crystal and couple
them to each other, such that entanglement and two-qubit gates would be
achieved.
\end{minipage}}

\section{Photons}

Photons can also make good qubits and they can carry quantum information over long distances hardly being affected by noise or decoherence. The qubit states can be encoded, for example, in the polarization of a single photon, and one-qubit gates can be easily realized with optical elements \cite{Kok07,OB07}. Unfortunately optical QC has a serious drawback: the difficulty in implementing two-qubit gates. Realizing the nonlinearity required for entangling two qubits is challenging, so alternatives such as the teleportation of nondeterministic quantum gates have been investigated \cite{OB07}. While this approach is still impractical due to the large amount of required resources, another solution  may be found in measurement-based QC.
\par
For the moment photons may not be practical as memory or computation qubits, but they are certainly the best ``flying qubits''.  Recent advances in quantum communication and, in particular, quantum key distribution are reviewed in \cite{Gis07}.

\section{Hybrids}

Exploiting the advantages of both natural and artificial atoms in hybrid systems provides exciting prospects for realizing QC. For instance, ions \cite{Tia04,Tia05} and atoms \cite{Ver08,Pet09} interfaced with superconducting circuits are now being investigated. As recent results point out neutral atoms and ions could also be interfaced with each other \cite{Zip10,Doe10}. While cavity QED with atoms and ions has been studied for some time now \cite{Kim08,Sho08}, solid-state cavity QED is more recent \cite{Sil07,Maj07,Eng07,Sho08}. For natural atoms strong coupling has been demonstrated \cite{Col07,Her09}. As mentioned before, in circuit QED the coupling strength is many orders of magnitude larger than in cavity QED, which is very promising for the study of quantum optics on a chip. As shown in Table 3, all systems discussed in the previous sections can be coupled with other systems. It is interesting to note that superconducting circuits can be coupled with both different types of  natural atoms, spins in solids \cite{Kub10,Sch10,Wu10} and with photons.
\par
Natural atoms, with their long decoherence times, are envisaged by many as quantum memories \cite{Sim10}, while the tunable artificial atoms may be used for the ``quantum processing unit". Both natural and artificial atoms may be coupled with photons via a cavity. Note that a necessary requirement is for the coupling timescale to be shorter than the decoherence time. Such cavities could be used as input/output interfaces and for long distance communication. Perhaps the first functional quantum computer will be a complex hybrid system made of natural atoms, artificial atoms, and photons.  Such a hybrid device is represented schematically in Figure 1. Several  types of hybrids are discussed in \cite{Wall09}.
\begin{figure}
\includegraphics[width=0.6\textwidth ]{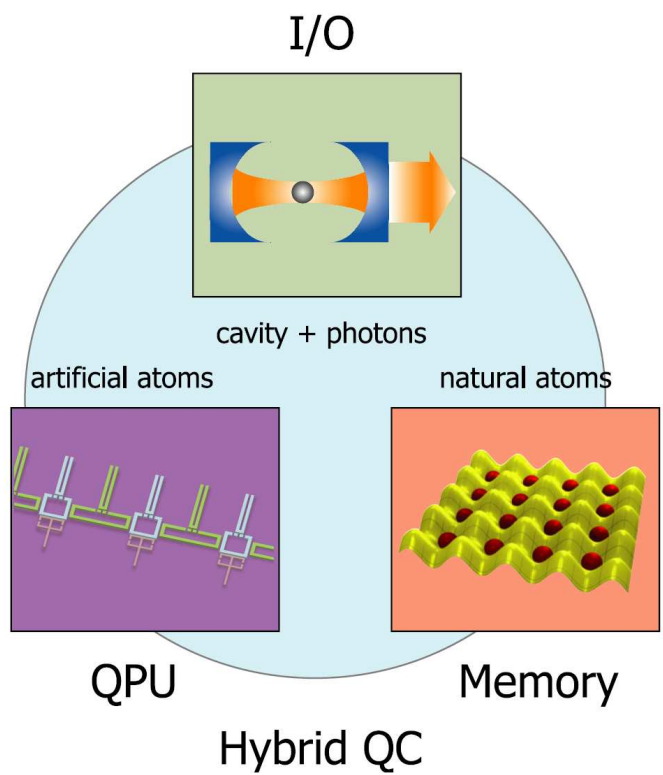}
\caption{Schematic representation of a hybrid device consisting of
natural atoms as quantum memory, artificial atoms as ``quantum
processing unit" (QPU), and an input/output (I/O) photonic
interface.}
\end{figure}

\section{Prospects}

In both natural and artificial atoms, almost all the basic requirements for realizing QC \cite{Div95}  have been demonstrated (i.e., (i) a scalable system with well-characterized qubits; (ii) initialization of the qubits; (iii) reasonably long decoherence times; (iv) a universal set of quantum gates; (v) measurement of the qubits).  Tables 1-6 and Figure 2 provide a brief snapshot of the progress and current experimental status for several types of qubits.
\par
The current challenges are to attain increased controllability (and minimize decoherence) and scale the existing systems to tens and hundreds of qubits and many-gate operations.  At this stage, new milestones, such as the creation of many-particle entangled states, the implementation of small quantum algorithms, and other applications (e.g., quantum simulation),  and the realization of quantum communication by interfacing the qubits with photons, are being targeted.
\par
``Quantum supercomputers'' for factorizing large numbers are still a distant goal. The first-generation of practical quantum computers  may be  either specialized devices for scientific applications like quantum simulations \cite{Bul09}, or integrated in complex quantum networks \cite{Kim08}.  As the very positive results summarized above point out, the first-generation quantum computers may be available in the near future. Furthermore, they may come as hybrids consisting of  natural atoms, artificial atoms, and photons.

\ack{
We thank R. Blatt, P. Grangier, L. Kouwenhoven, C. Marcus, A. Morello, W. Oliver, T. Porto, M. Saffman, D. Wineland and A. Yacoby for useful comments on the manuscript.
\\
FN acknowledges partial support from the
Laboratory of Physical Sciences (LPS),
National Security Agency (NSA), Army Research Office (ARO),
Defense Advanced Research Projects Agency (DARPA),
Air Force Office of Scientific Research (AFOSR),
National Science Foundation (NSF) grant No.0726909,
JSPS-RFBR contract No.09-02-92114,
Grant-in-Aid for Scientific Research (S),
MEXT Kakenhi on Quantum Cybernetics, and the
Funding Program for Innovative R$\&$D on Science and Technology (FIRST).
}

\begin{figure}[h!]
\includegraphics[width=0.8\textwidth ]{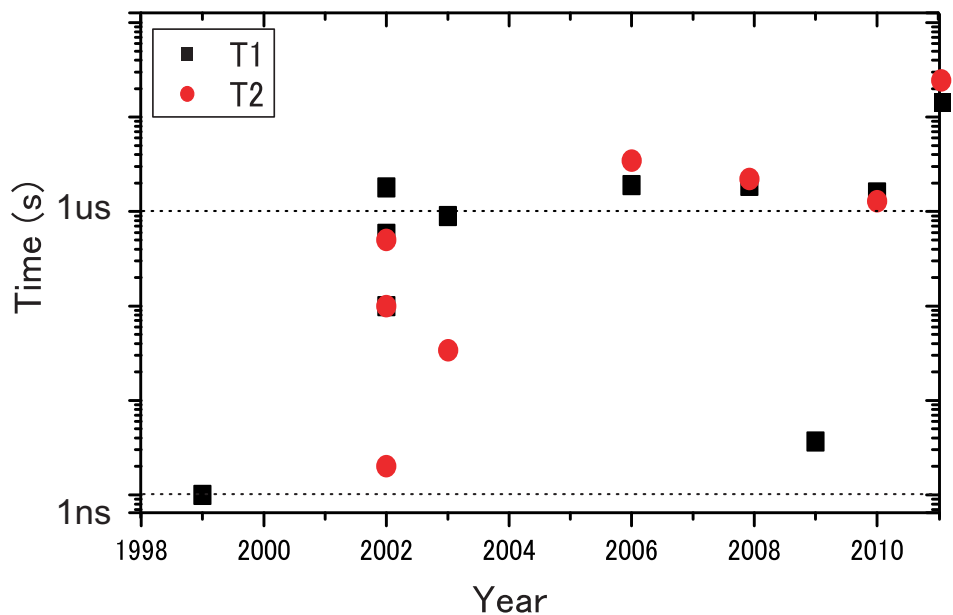}
\caption{An example of the progress that has been achieved for superconducting circuits
in the last decade. The decoherence time kept increasing, and the current trend
promises decoherence times of the order of ms in the next couple of years. Visibility
also increased and now it is larger than $95\%$. The black squares show $T_1$ and the red dots $T_2$.}
\end{figure}

\begin{table}[h!]
\caption{\label{tabl1}Comparison between natural and artificial atoms.Note: $^{\rm{(a)}}$ distance between qubits for NV centers and $^{\rm{(b)}}$ typical distances between quantum dots.\vspace{8pt}}
\small
\begin{tabular}{@{}lcccc}
\br
&\centre{2}{\textbf{Natural atoms}}&\centre{2}{\textbf{Artificial atoms}}\\
\ns
 &Neutral atoms&Trapped ions&Supercond. circuits&Spins in solids\\
\mr
\textbf{Energy gap}&GHz (hyperfine),&GHz (hyperfine),&1 $-$ 10 GHz& GHz,\\
&$10^{14}$ Hz (optical)&$10^{14}$ Hz (optical)&& $10^{13}$ Hz\\[8pt]
\textbf{Photon}&Optical, MW&Optical, MW&MW&Optical, MW,\\
&&&&infrared\\[8pt]
\textbf{Dimension}&$\sim 2$ $\rm{\AA}$&$\sim 2$ $\rm{\AA}$&$\sim \mu$m&$\sim$ nm\\[8pt]
\textbf{Distance}&$<1$ $\mu$m&$\sim 5$ $\mu$m&$\sim \mu$m&$\sim 10$ nm $^{\rm{(a)}}$, $\sim 100$ nm $^{\rm{(b)}}$\\
\textbf{between qubits}&&&&\\[8pt]
\textbf{Operating}&nK$-$ $\mu$K&$\mu$K $-$ mK&$\sim$ mK&mK $-$ 300 K\\
\textbf{temperature}&&&&\\[8pt]
\textbf{Qubit}&Collisions,&Coulomb&Capacitive,&Coulomb,\\
\textbf{interactions}&exchange&&inductive&exchange,\\
&&&&dipolar\\[8pt]
\textbf{Cooling}&Doppler,&Doppler,&Cryogenic&Cryogenic\\
&Sisyphus,&sideband&&\\
&evaporative&&&\\[8pt]
\textbf{Cavity}&Optical,&Optical,&Transmission&Optical,\\
&MW&vib. modes&line, LC&MW\\
&&&circuit&\\
\br
\end{tabular}

\end{table}

\begin{table}[h!]
\caption{\label{tabl2}Comparison between natural and artificial atoms in the view of implementing quantum computation. Hereafter, MW stands for microwaves and SC for superconducting. $^{\rm {(a)}}$ large entangled states can also be realized with collisional gates; $^{\rm{(b)}}$ entanglement of the ground state of four qubits; $^{\rm{(c)}}$ NV centers in diamond; $^{\rm{(d)}}$ $T_1$ for the vibrational modes; $^{\rm{(e)}}$ $T_1$ for the internal hyperfine states; $^{\rm{(f)}}$ of the order of ms for NV centers at room temperature and of the order of minutes at 1 K; of the order of seconds for P:Si;  $^{\rm{(g)}}$ in optical clocks $T_1$, $T_2>10$ minutes has been observed; $^{\rm{(i)}}$ only generated for one and two resonators and not for many qubits.\vspace{8pt}}
\small
\begin{tabular}{@{}lcccc}
\br
&\centre{2}{\textbf{Natural atoms}}&\centre{2}{\textbf{Artificial atoms}}\\
\ns
 &Neutral atoms&Trapped ions&Supercond. circuits&Spins in solids\\
\mr
\textbf{$\#$ entangled qubits}&2 $^{\rm{(a)}}$&14&3 (4 $^{\rm{(b)}}$)&1 (3 $^{\rm{(c)}}$)\\[8pt]
\textbf{One-qubit gates fidelity}&$99\%$&$99\%$& $99\%$&$>73\%$ ($>99\%$ $^{\rm{(c)}}$)\\[8pt]
\textbf{Two-qubit gates fidelity}&$>64\%$&$99.3\%$& $>90\%$& $90\%$ $^{\rm{(c)}}$\\[8pt]
\textbf{Entangled states}&Bell&Bell, GHZ,& Bell, GHZ $^{\rm{(i)}}$&GHZ $^{\rm{(c)}}$\\
&&W, cat&W, cat& \\[8pt]
\textbf{Measurement efficiency}&$99.9\%$&$99.9\%$& $>95\%$&$99\%$\\[8pt]
\textbf{$T_1$}&$\sim$ s&$\sim 100$ ms $^{\rm{(d)}}$& 10 $\mu$s&$\sim 1$ s $^{\rm{(f)}}$\\
&&$>20$ ms $^{\rm{(e)}}$ && \\[8pt]
\textbf{$T_2$}&$\sim40$ ms&1000 s $^{\rm{(g)}}$& 20 $\mu$s&200 $\mu$s $^{\rm{(f)}}$\\[8pt]
\textbf{$Q_1$}&$\sim10^4$&$\sim10^{13}$& $\sim10^5$&$\sim10^3$ $-$ $10^4$\\
& & & & ($10^6$ $^{\rm{(c)}}$) \\[8pt]
\textbf{$Q_2$}&$\sim 4\times 10^4$&$2 \times 10^2$ $-$ $2\times 10^3$& $>100$&tbd\\
&&$\sim 2\times 10^4$&& \\[8pt]
\textbf{Interfaceable with}&photons, SC&photons, SC& photons, atoms,&photons\\
&circuits&circuits&ions& \\
\br
\end{tabular}
\end{table}
\clearpage
\begin{table}[h!]
\caption{\label{label}Interfacing different types of qubits for future scalability or realizing long-range quantum communication. The asterisk denotes the cases that have been experimentally realized and the dash means that, to the best of our knowledge, no proposal exists yet.\vspace{8pt}}
\large
\begin{tabular}{@{}lccccc}
\br
&\textbf{Atoms}&\textbf{Ions}&\textbf{Cavity}&\textbf{Spins}&\textbf{SC}\\
\mr
\textbf{Atoms}& &\checkmark &\checkmark $*$&-&\checkmark $*$\\ [6pt]
\textbf{Ions}&\checkmark &&\checkmark $*$&-&\checkmark\\ [6pt]
\textbf{Cavity}&\checkmark $*$&\checkmark $*$&&\checkmark& \checkmark $*$\\ [6pt]
\textbf{Spins}&-&-&\checkmark & & \checkmark \\[6pt]
\textbf{SC}&\checkmark $*$&\checkmark &\checkmark $*$ &\checkmark & \\
\br
\end{tabular}

\end{table}

\begin{table}[h!]
\caption{\label{tabl4}Coherence times of superconducting qubits.\vspace{8pt}}
\begin{tabular}{@{}lcccc}
\br
\textbf{Year}&\textbf{T1}&\textbf{T2 (echo)}&\textbf{Qubit}&\textbf{Ref.}\\
\mr
\textbf{1999}&1 ns & $-$ & Charge&\cite{Nak99}\\[6pt]
\textbf{2002}&580 ns & 2 ns & Charge&\cite{Nak02}\\[6pt]
\textbf{2002}&100 ns & 100 ns& Phase&\cite{Mar02}\\[6pt]
\textbf{2002}&1.8 $\mu$s & 500 ns & Hybrid (charge/phase) &\cite{Vio02}\\[6pt]
\textbf{2003}&0.9 $\mu$s & 30 ns & Flux &\cite{Chi03}\\[6pt]
\textbf{2006}&1.9 $\mu$s & 3.5 $\mu$s & Flux &\cite{Yos06}\\[6pt]
\textbf{2008}&1.87 $\mu$s & 2.22 $\mu$s & Hybrid (charge/phase) &\cite{Sch08}\\[6pt]
\textbf{2009}&350 ns & $-$ & Flux &\cite{Man09}\\[6pt]
\textbf{2010}&1.6 $\mu$s & 1.3 $\mu$s & Hybrid (phase/flux) &\cite{Ste10}\\[6pt]
\textbf{2011}&12 $\mu$s & 23 $\mu$s & Flux &\cite{Byl11}\\[6pt]
\textbf{2011}&0.2 ms & - & Charge &\cite{Kim11}\\[6pt]
\br
\end{tabular}
\end{table}

\begin{table}[h!]
\caption{\label{tabl5}Progress in the implementation of superconducting qubits quantum gates.\vspace{8pt}}
\small
\begin{tabular}{@{}lcccc}
\br
\textbf{Year}&\textbf{Operation}&\textbf{Qubits}&\textbf{Mechanism}&\textbf{Ref.}\\
\mr
\textbf{2003}&CNOT gate & 2 & Direct coupling; &\cite{Yam03}\\
&&&gate relies on zz component &\\[6pt]
\textbf{2003}&Entangled energy levels & 2 & Direct xy coupling&\cite{Ber03}\\[6pt]
\textbf{2005}&iSWAP; Entanglement & 2& Direct xy coupling&\cite{McD05}\\[6pt]
\textbf{2006}&iSWAP; Entanglement & 2 & Direct xy coupling &\cite{Ste06}\\[6pt]
\textbf{2006}&Entangled energy levels& 4& Direct coupling &\cite{Gra06}\\[6pt]
\textbf{2006-7}&Controllable coupling& 2 & Coupling mediated by &\cite{Him06,Har07}\\
&& & additional circuit element &\\[6pt]
\textbf{2007}&CNOT gate& 2 & Direct coupling;&\cite{Pla07}\\
&& & gate relies on zz component &\\[6pt]
\textbf{2007}&iSWAP & 2 & xy coupling to cavity; &\cite{Sil07}\\
& & &gate mediated by cavity &\\[6pt]
\textbf{2007}&iSWAP & 2 & xy coupling mediated by cavity &\cite{Maj07}\\[6pt]
\textbf{2007}&iSWAP & 2 & Coupling mediated by additional  &\cite{Nis07}\\
& & & circuit element; gate relies on xy coupling &\\[6pt]
\textbf{2009}&CPhase & 2 & zz coupling mediated by &\cite{DiC09}\\
& &  & auxilliary energy levels &\\[6pt]
\textbf{2010}&Entanglement & 3 & xy coupling &\cite{Nee10}\\[6pt]
\textbf{2010}&Entanglement & 3 & zz coupling mediated by &\cite{diC10}\\
& &  & auxilliary energy levels &\\[6pt]
\textbf{2011}&3-qubit gate & 3 & Coupling mediated by &\cite{Mar11}\\
& &  & auxilliary energy levels &\\
\br
\end{tabular}

\end{table}

\begin{table}[h!]
\caption{\label{tabl6}Progress in the number of qubits and fidelities for different operations on trapped ions. CZ stands for the Cirac-Zoller scheme \cite{Cir95}, and MS for the M$\rm{\o}$lmer-S$\rm{\o}$rensen scheme \cite{Mol99}.\vspace{8pt}}
\small
\begin{tabular}{@{}lccccc}
\br
\textbf{Year}&\textbf{Operation}&\textbf{Mechanism}&\textbf{Qubits}&\textbf{Fidelity}&\textbf{Ref.}\\
\mr
\textbf{1998}&Entanglement & CZ & 2&70$\%$ &\cite{Tur98}\\[6pt]
\textbf{2000}&Entanglement & MS & 2&83$\%$ &\cite{Sac00}\\
& &  & 4&57$\%$ &\\[6pt]
\textbf{2003}&CNOT gate & CZ & 2&71.3$\%$ &\cite{Sch03}\\[6pt]
\textbf{2003}&Entanglement & Geometric & 2&97$\%$ &\cite{Lei03}\\[6pt]
\textbf{2005}&Entanglement & CZ & 4&$>$76$\%$ &\cite{Lei05}\\
& &  & 5&$>$60$\%$ &\\
& &  & 6&$>$50$\%$ &\\[6pt]
\textbf{2005}&Entanglement & CZ & 4&85$\%$ &\cite{Haf05}\\
& &  & 5&76$\%$ &\\
& &  & 6&79$\%$ &\\
& &  & 7&76$\%$ &\\
& &  & 8&72$\%$ &\\[6pt]
\textbf{2006}&CNOT gate & CZ & 2&92.6$\%$ &\cite{Rie06}\\[6pt]
\textbf{2008}&Entanglement & MS & 2&99.3$\%$ &\cite{Ben08}\\[6pt]
\textbf{2009}&Toffoli gate & CZ & 3&74$\%$ &\cite{Mon09}\\[6pt]
\textbf{2010}&Entanglement & MS & 10&62.9$\%$ &\cite{Mon10}\\
& &  & 12&39.6$\%$ &\\
& &  & 14&46.3$\%$ &\\
\br
\end{tabular}
\end{table}

\clearpage
\appendix
\section{Tables summarizing the main characteristics of different systems in the view of realizing QC}

In the following tables, $T_1$ (relaxation time) is defined as the average time that the system takes for its excited state to decay to the ground state; $T_2$ (decoherence time) represents the average time over which the qubit energy-level difference does not vary; $Q_1$ (quality factor) represents the number of one-qubit quantum gates that can be realized within the time $T_2$; $Q_2$ (quality factor) represents the number of two-qubit quantum gates that can be realized within the time $T_2$. The following abbreviation is used: tbd for ``to be demonstrated''
\clearpage
\begin{table}[hb]
\caption{\label{at}Neutral atoms.\vspace{5pt}}
\footnotesize
\begin{tabular}{@{}ll}
\br
&{\textbf{Neutral atoms}}\\[6pt]
\ns
\mr
Qubits&Internal states (ground hyperfine states);\\
&Motional states (trapping potential eigenstates)\\[6pt]
Scalability&Demonstrated in optical lattices; possible in arrays of cavities, atom chips\\[6pt]
Initialization&Both internal (optical pumping) and motional (laser cooling) states\\[6pt]
Long coherence time&Several seconds \cite{Yav06,Deu10,Tre04}\\[6pt]
Universal quantum gates&One-, two-qubit gates (several proposals)\\[6pt]
Measurement&Fluorescence:``quantum jump'' technique\\[6pt]
\br
\textbf{Fabrication}&\\[6pt]
\ns
\mr
Material&Trapped neutral atoms: Rb, Li, K, Cs, etc\\[6pt]
Well controlled fabrication&yes\\[6pt]
Flexible geometry&yes (especially in optical lattices)\\[6pt]
Distance between qubits&A few hundred nm to a few $\mu$m \cite{Bloch08}\\[6pt]
\br
\textbf{Operation}&\\[6pt]
\ns
\mr
Qubits demonstrated&$>10^6$ (stored), 2 (entangled)\\[6pt]
Superposition/Entangled states&yes/yes\\[6pt]
One-qubit gates (Fidelity)&yes (99.98 $\%$)\\[6pt]
Two-qubit gates (Fidelity)&yes (SWAP $>$64$\%$ \cite{And07}); CNOT (73$\%$ \cite{Ise10})\\[6pt]
Operation temperature&From nK to$\mu$K\\
\br
\textbf{Readout}&\\[6pt]
\ns
\mr
Readout (Fidelity) &Laser-induced fluorescence ($99.9\%$)\\[6pt]
Single-qubit readout possible&yes\\
\br
\textbf{Manipulation}&\\[6pt]
\ns
\mr
Controls &Optical fields, microwave\\[6pt]
Types of operations &One-, two-qubit gates, entanglement\\[6pt]
Individual addressing&To be demonstrated \cite{Nel07,Wur09,She10,Gib10,Fuh10}\\
\br
\textbf{Decoherence}&\\[6pt]
\ns
\mr
Decoherence sources &Photon scattering, heating, stray fields, laser fluctuations\\[6pt]
$T_1$ &$\sim$ s\\[6pt]
$T_2$ & $\sim$ 40 ms\\[6pt]
$Q_1$ & $\sim 10^4$\\[6pt]
$Q_2$ &$\sim$ 40000\\
\br
\end{tabular}
\end{table}

\begin{table}[hb]
\caption{\label{ions}Trapped ions.\vspace{5pt}}
\footnotesize
\begin{tabular}{@{}ll}
\br
&{\textbf{Trapped ions}}\\[6pt]
\ns
\mr
Qubits&Internal states (hyperfine or Zeeman sublevels, optical);\\
&Motional states (collective oscillations)\\[6pt]
Scalability&Ion shuttling, arrays, photon interconnections, long strings\\[6pt]
Initialization&Both internal (optical pumping) and motional (laser cooling) states\\[6pt]
Long coherence time&Internal: hyperfine $>20$ s, optical $>1$ s; Motional: $\sim 100$ ms\\[6pt]
Universal quantum gates&One-, two-, three-qubit gates\\[6pt]
Measurement&Fluorescence:``quantum jump'' technique\\[6pt]
\br
\textbf{Fabrication}&\\[6pt]
\ns
\mr
Material&Atomic ions: Ca$^+$, Be$^+$, Ba$^+$, Mg$^+$, etc\\[6pt]
Well controlled fabrication&yes\\[6pt]
Flexible geometry&yes\\[6pt]
Distance between qubits&A few $\mu$m to tens of $\mu$m\\[6pt]
\br
\textbf{Operation}&\\[6pt]
\ns
\mr
Qubits demonstrated&$10-10^3$ (stored), 14 (entangled) \cite{Mon10}\\[6pt]
Superposition/Entangled states&yes/yes (2-14 ions, fidelities $99.3\%$-$46\%$) \cite{Mon10}\\[6pt]
One-qubit gates (Fidelity)&yes ($99\%$)\\[6pt]
Two-qubit gates (Fidelity)&yes (CNOT $>99.3\%$ \cite{Ben08}; Toffoli $71.3\%$ \cite{Mon09}; gate time 1.5 ms)\\[6pt]
Operation temperature&From $\mu$K to mK\\
\br
\textbf{Readout}&\\[6pt]
\ns
\mr
Readout (Fidelity) &Laser-induced fluorescence (99.9$\%$)\\[6pt]
Single-qubit readout possible&yes\\
\br
\textbf{Manipulation}&\\[6pt]
\ns
\mr
Controls &Optical, microwave, electric/magnetic fields\\[6pt]
Types of operations &One-, two-, three-qubit gates, entanglement\\[6pt]
Individual addressing&yes\\
\br
\textbf{Decoherence}&\\[6pt]
\ns
\mr
Decoherence sources &Heating, spontaneous emission, laser, magnetic field fluctuations\\[6pt]
$T_1$ & a few minutes (hyperfine), 1 s (optical), 100 ms (motional)\\[6pt]
$T_2$ & 15 s\\[6pt]
$Q_1$ & $\sim10^{13}$ (single-qubit gate 50 ps) \cite{Cam10}\\[6pt]
$Q_2$ &$\sim 20000$ (MS gate 50 $\mu$s) \cite{Ben08}; $\sim 200$ (CZ gate 500 $\mu$s) \cite{Rie06}\\
\br

\end{tabular}
\end{table}
\begin{table}[hb]
\caption{\label{nmr}Nuclear spins manipulated by Nuclear Magnetic Resonance (NMR).\vspace{5pt}}
\footnotesize
\begin{tabular}{@{}ll}
\br
&{\textbf{NMR}}\\[6pt]
\ns
\mr
Qubits&Nuclear spin\\[6pt]
Scalability&Not available in liquid-state NMR; possible for solid-state NMR\\[6pt]
Initialization&Demonstrated\\[6pt]
Long coherence time&$>1$ s\\[6pt]
Universal quantum gates&One-, two-, three-qubit gates\\[6pt]
Measurement&Single-qubit measurement not available\\[6pt]
\br
\textbf{Fabrication}&\\[6pt]
\ns
\mr
Material&Organic molecules (alanine, chloroform, cytosine)\\[6pt]
Well controlled fabrication&yes\\[6pt]
Flexible geometry&no\\[6pt]
Distance between qubits&$\sim \rm{\AA}$\\[6pt]
\br
\textbf{Operation}&\\[6pt]
\ns
\mr
Qubits demonstrated&7, 12 (entangled) liquid-state \cite{Neg06}; $>$100 (correlated) solid-state\\[6pt]
Superposition/Entangled states&yes/yes \\[6pt]
One-qubit gates (Fidelity)&yes ($>98\%$)\\[6pt]
Two-qubit gates (Fidelity)&yes ($>98\%$ CNOT and SWAP)\\[6pt]
Operation temperature&Room temperature\\
\br
\textbf{Readout}&\\[6pt]
\ns
\mr
Readout (Fidelity) &Voltage in neighboring coil induced by precessing spins, $99.9\%$\\[6pt]
Single-qubit readout possible&no\\
\br
\textbf{Manipulation}&\\[6pt]
\ns
\mr
Controls &RF pulses\\[6pt]
Types of operations &One-, two-, three-qubit gates\\[6pt]
Individual addressing&no\\
\br
\textbf{Decoherence}&\\[6pt]
\ns
\mr
Decoherence sources &Coupling errors\\[6pt]
$T_1$ &$>1$ s (liquid-state); $>1$ min (solid-state)\\[6pt]
$T_2$ & $\sim 1$ s (liquid-state); $>1$ s (solid-state)\\[6pt]
$Q_1$ &\\[6pt]
$Q_2$ &100 (gate time 10 ms)\\
\br

\end{tabular}
\end{table}
\begin{table}[hb]
\caption{\label{sc}Superconducting circuits.\vspace{5pt}}
\footnotesize
\begin{tabular}{@{}ll}
\br
&{\textbf{Superconducting circuits}}\\[6pt]
\ns
\mr
Qubits&Flux, phase states, charge; also hybrids\\[6pt]
Scalability&High potential for scalability\\[6pt]
Initialization&Demonstrated for all types of qubits\\[6pt]
Long coherence time&$\sim 10$ $\mu$s\\[6pt]
Universal quantum gates&One-, two-qubit gates\\[6pt]
Measurement&Individual measurement possible\\[6pt]
\br
\textbf{Fabrication}&\\[6pt]
\ns
\mr
Material&Josephson junctions (Al-Al$_x$O$_y$-Al,Nb-Al$_x$O$_y$Nb)\\[6pt]
Well controlled fabrication&yes\\[6pt]
Flexible geometry&yes\\[6pt]
Distance between qubits&$\sim \mu$m\\[6pt]
\br
\textbf{Operation}&\\[6pt]
\ns
\mr
Qubits demonstrated&128 (fabricated) \cite{Har10}, 3 (entangled)\\[6pt]
Superposition/Entangled states&yes/yes \\[6pt]
One-qubit gates (Fidelity)&yes ($99\%$)\\[6pt]
Two-qubit gates (Fidelity)&yes ($>90\%$) \cite{DiC09}\\[6pt]
Operation temperature&mK\\
\br
\textbf{Readout}&\\[6pt]
\ns
\mr
Readout (Fidelity) &SET, SQUID ($>95\%$) \cite{Pic08}, cavity frequency shift \cite{Wal04}\\[6pt]
Single-qubit readout possible&yes\\
\br
\textbf{Manipulation}&\\[6pt]
\ns
\mr
Controls &Microwave pulses, voltages, currents\\[6pt]
Types of operations &One-, two-, three-qubit gates, entanglement\\[6pt]
Individual addressing&yes\\
\br
\textbf{Decoherence}&\\[6pt]
\ns
\mr
Decoherence sources &Electric and magnetic noise, 1/f noise\\[6pt]
$T_1$ & 0.2 ms \cite{Kim11}\\[6pt]
$T_2$ & 23 $\mu$s \cite{Byl11}\\[6pt]
$Q_1$ &$\sim 10^5$\\[6pt]
$Q_2$ &$>100$ (gate time 10-50 ns) \cite{DiC09}\\
\br

\end{tabular}
\end{table}
\begin{table}[hb]
\caption{\label{spi}Spins in solids. Here, QDs stand for quantum dots, NV centers for nitrogen-vacancy centers in diamond and P:Si for phosphorous on silicon. The asterisk $*$ refers to room temperature. \vspace{5pt}}
\footnotesize
\begin{tabular}{@{}ll}
\br
&{\textbf{Spins in solids}}\\[6pt]
\ns
\mr
Qubits&Electron spin; Electron and nuclear spins in NV centers in diamond, P:Si\\[6pt]
Scalability&High potential for scalability\\[6pt]
Initialization&Demonstrated \\[6pt]
Long coherence time&$>1$ s (QDs); $\sim$ s (NV centers), $\sim 100$ s (P:Si)\\[6pt]
Universal quantum gates&One-qubit gates\\[6pt]
Measurement&Electrical, optical\\[6pt]
\br
\textbf{Fabrication}&\\[6pt]
\ns
\mr
Material&GaAs, InGaAs (QDs), NV centers in diamond, P:Si\\[6pt]
Well controlled fabrication&yes\\[6pt]
Flexible geometry&yes\\[6pt]
Distance between qubits&100-300 nm(QDs); $\sim10$ nm (NV centers)\\[6pt]
\br
\textbf{Operation}&\\[6pt]
\ns
\mr
Qubits demonstrated&1 (QDs), 3 (NV centers) \cite{Neu08}\\[6pt]
Superposition&yes \\[6pt]
One-qubit gates (Fidelity)&yes ($>73\%$ QDs \cite{Kop06}; $>99\%$ NV centers \cite{Lan10})\\[6pt]
Two-qubit gates (Fidelity)&yes ($90\%$ NV centers \cite{Jele04})\\[6pt]
Operation temperature&From mK to a few K (QDs); room temperature (NV centers)\\
\br
\textbf{Readout}&\\[6pt]
\ns
\mr
Readout (Fidelity) &electrical, optical (90-92$\%$)\\[6pt]
Single-qubit readout possible&yes\\
\br
\textbf{Manipulation}&\\[6pt]
\ns
\mr
Controls &RF, optical pulses, electrical\\[6pt]
Types of operations &One-qubit gates ($>$73$\%$ gate time 25 ns)\\[6pt]
Individual addressing&yes\\
\br
\textbf{Decoherence}&\\[6pt]
\ns
\mr
Decoherence sources &Co-tunneling, charge noise, coupling with nuclear spins\\[6pt]
$T_1$ &$>1$ s (QDs) \cite{Ama08}; $>5$ ms $^{*}$ (NV centers) \cite{Neu08}; 6 s \cite{Mor10} (P:Si); 100 s \cite{MCC10} (P:Si)\\[6pt]
$T_2$ & $\sim 270$ $\mu$s \cite{Blu10,Bar10}; $\sim 1.8$ ms (NV centers) \cite{Bal09}; $\sim 60$ ms \cite{Tyr03} (P:Si); 2 s \cite{Mor08} (P:Si) \\[6pt]
$Q_1$ &$\sim 10^3$ (gate time 180 ps); $\sim 10^4$ (gate time 30 ps) \cite{Ber08}; $>10^6$ (gate time $\sim 1$ ns)\\[6pt]
$Q_2$ &tbd\\
\br

\end{tabular}
\end{table}

\clearpage
\section*{References}

Due to space limitations we list a small subset of recent, relevant papers, mostly experimental results. The very few theory papers cited here introduce parameters used in the experimental papers cited, and also in the tables (e.g., as in Table \ref{tabl6}). For more references on the theoretical aspects, please refer to the various more specialized reviews listed below.\vspace{8pt}
\\
\\
\bibliography{ROPqubits}

\begin{thebibliography}{100}

\bibitem{Bloch08}
I.~Bloch.
\newblock Quantum coherence and entanglement with ultracold atoms in optical
  lattices.
\newblock {\em Nature}, 453:1016--1022, 2008.

\bibitem{Bla08}
R.~Blatt and D.~J. Wineland.
\newblock Entangled states of trapped atomic ions.
\newblock {\em Nature}, 453:1008--1015, 2008.

\bibitem{CW08}
J.~Clarke and F.~K. Wilhelm.
\newblock Superconducting quantum bits.
\newblock {\em Nature}, 453:1031--1042, 2008.

\bibitem{You05}
J.~Q. You and F.~Nori.
\newblock Superconducting circuits and quantum information.
\newblock {\em Physics Today}, 58(11):42--47, 2005.

\bibitem{HA08}
R.~Hanson and D.~D. Awschalom.
\newblock Coherent manipulation of single spins in semiconductors.
\newblock {\em Nature}, 453:1043--1049, 2008.

\bibitem{VC05}
L.~M.~K. Vandersypen and I.~L. Chuang.
\newblock {NMR} techniques for quantum control and computation.
\newblock {\em Rev. Mod. Phys.}, 76:1037--1069, 2005.

\bibitem{Bau07}
J.~Baugh, J.~Chamilliard, C.~M. Chandrashekar, M.~Ditty, A.~Hubbard,
  R.~Laflamme, M.~Laforest, D.~Maslov, O.~Moussa, C.~Negrevergne, M.~Silva,
  S.~Simmons, C.~A. Ryan, D.~G. Cory, J.~S. Hodges, and C.~Ramanathan.
\newblock Quantum information processing using nuclear and electron magnetic
  resonance: review and prospects.
\newblock {\em arXiv:0710.1447v1}, 2007.

\bibitem{Kan98}
B.~E. Kane.
\newblock A silicon-based nuclear spin quantum computer.
\newblock {\em Nature}, 393:133--137, 1998.

\bibitem{Mor08}
J.~J.~L. Morton, A.~M. Tyryshkin, R.~M. Brown, S.~Shankar, B.~W. Lovett,
  A.~Ardavan, T.~Schenkel, E.~E. Haller, J.~W. Ager, and S.~A. Lyon.
\newblock Solid-state quantum memory using the 31{P} nuclear spin.
\newblock {\em Nature}, 455:1085--1088, 2008.

\bibitem{Gis07}
N.~Gisin and R.~Thew.
\newblock Quantum communication.
\newblock {\em Nature Photonics}, 1:165--171, 2007.

\bibitem{Kok07}
P.~Kok, W.~J. Munro, K.~Nemoto, T.~C. Ralph, J.~P. Dowling, and G.~J. Milburn.
\newblock Linear optical quantum computing with photonic qubits.
\newblock {\em Reviews of Modern Physics}, 79:135, 2007.

\bibitem{Lad10}
T.~D. Ladd, F.~Jelezko, Y.~Nakamura R.~Laflamme~and, C.~Monroe, and J.~L.
  O'Brien.
\newblock Quantum computers.
\newblock {\em Nature}, 464:45--53, 2010.

\bibitem{Man03}
O.~Mandel, M.~Greiner, A.~Widera, T.~Rom, T.~W. H\"ansch, and I.~Bloch.
\newblock Controlled collisions for multi--particle entanglement of optically
  trapped atoms.
\newblock {\em Nature}, 425:937--940, 2003.

\bibitem{Sch04}
D.~Schrader, I.~Dotsenko, M.~Khudaverdyan, Y.~Miroshnychenko,
  A.~Rauschenbeutel, and D.~Meschede.
\newblock Neutral atom quantum register.
\newblock {\em Phys. Rev. Lett.}, 93:150501, 2004.

\bibitem{Tre04}
P.~Treutlein, T.~W.~H\"ansch P.~Hommelhoff, T.~Steinmetz, and J.~Reichel.
\newblock Coherence in microchip traps.
\newblock {\em Phys. Rev. Lett.}, 92:203005, 2004.

\bibitem{Jak05}
D.~Jaksch and P.~Zoller.
\newblock The cold atom {H}ubbard toolbox.
\newblock {\em Annals of Physics}, 315:52--79, 2005.

\bibitem{Mic06}
A.~Micheli, G.~K. Brennen, and P.~Zoller.
\newblock A toolbox for lattice--spin models with polar molecules.
\newblock {\em Nature Physics}, 2:341--347, 2006.

\bibitem{Mir06}
Y.~Miroshnychenko, W.~Alt, I.~Dotsenko, L.~Forster, M.~Khudaverdyan,
  D.~Meschede, D.~Schrader, and A.~Rauschenbeutel.
\newblock Quantum engineering: An atom-sorting machine.
\newblock {\em Nature}, 442:151, 2006.

\bibitem{Yav06}
D.~D. Yavuz, P.~B. Kulatunga, E.~Urban, T.~A. Johnson, N.~Proite, T.~Henage,
  T.~G. Walker, and M.~Saffman.
\newblock Fast ground state manipulation of neutral atoms in microscopic
  optical traps.
\newblock {\em Phys. Rev. Lett.}, 96(6):063001, 2006.

\bibitem{And07}
M.~Anderlini, B.~L.~Brown P.~J. Lee~and, J.~Sebby-Strabley, W.~D. Phillips, and
  J.~V. Porto.
\newblock Controlled exchange interaction between pairs of neutral atoms in an
  optical lattice.
\newblock {\em Nature}, 448:452--456, 2007.

\bibitem{Beu07}
J.~Beugnon, C.~Tuchendler, H.~Marion, A.~Gaetan, Y.~Miroshnychenko, Y.~R.~P.
  Sortais, A.~M. Lance, M.~P.~A. Jones, G.~Messin, A.~Browaeys, and
  P.~Grangier.
\newblock Two-dimensional transport and transfer of a single atomic qubit in
  optical tweezers.
\newblock {\em Nature Physics}, 3:696--699, 2007.

\bibitem{Hay07}
D.~Hayes, P.~S. Julienne, and I.~H. Deutsch.
\newblock Quantum logic via the exchange blockade in ultracold collisions.
\newblock {\em Phys. Rev. Lett.}, 98:070501, 2007.

\bibitem{Lew07}
M.~Lewenstein, A.~Sanpera, V.~Ahufinger, B.~Damski, A.~Sen(De), and U.~Sen.
\newblock Ultracold atomic gases in optical lattices: mimicking condensed
  matter physics and beyond.
\newblock {\em Advances in Physics}, 56:243--379, 2007.

\bibitem{Nel07}
K.~D. Nelson, X.~Li, and D.~S. Weiss.
\newblock Imaging single atoms in a three--dimensional array.
\newblock {\em Nature Physics}, 3:556--560, 2007.

\bibitem{Tro08}
S.~Trotzky, P.~Cheinet, S.~Folling, M.~Feld, U.~Schnorrberger, A.~M. Rey,
  A.~Polkovnikov, E.~A. Demler, M.~D. Lukin, and I.~Bloch.
\newblock Time-resolved observation and control of superexchange interactions
  with ultracold atoms in optical lattices.
\newblock {\em Science}, 319:295--299, 2008.

\bibitem{Gae09}
A.~Gaetan, Y.~Miroshnychenko, T.~Wilk, A.~Chotia, M.~Viteau, D.~Comparat,
  P.~Pillet, A.~Browaeys, and P.~Grangier.
\newblock Observation of collective excitation of two individual atoms in the
  {R}ydberg blockade regime.
\newblock {\em Nature Physics}, 5:115--118, 2009.

\bibitem{Saff09}
M.~Saffman, T.~G. Walker, and K.~M$\rm{\o}$lmer.
\newblock Quantum information with {R}ydberg atoms.
\newblock {\em Rev. Mod. Phys.}, 82:2313, 2010.

\bibitem{Urb09}
E.~Urban, T.~A. Johnson, T.~Henage, L.~Isenhower, D.~D. Yavuz, T.~G. Walker,
  and M.~Saffman.
\newblock Observation of {R}ydberg blockade between two atoms.
\newblock {\em Nature Physics}, 5:110--114, 2009.

\bibitem{Wur09}
P.~W\"urtz, T.~Langen, T.~Gericke, A.~Koglbauer, and H.~Otto.
\newblock Experimental demonstration of single--site addressability in a
  two-dimensional optical lattice.
\newblock {\em Phys. Rev. Lett.}, 103:080404, 2009.

\bibitem{Deu10}
C.~Deutsch, F.~Ramirez-Martinez, C.~Lacro\^ute, F.~Reinhard, T.~Schneider,
  J.~N. Fuchs, F.~Pi\'echon, F.~Lalo\"e, J.~Reichel, and P.~Rosenbusch.
\newblock Spin self-rephasing and very long coherence times in a trapped atomic
  ensemble.
\newblock {\em Phys. Rev. Lett.}, 105:020401, 2010.

\bibitem{Fuh10}
A.~Fuhrmanek, R.~Bourgain, Y.~R.~P. Sortais, and A.~Browaeys.
\newblock Free-space lossless state detection of a single trapped atom.
\newblock {\em Phys. Rev. Lett.}, 106:133003, 2011.

\bibitem{Gib10}
Michael~J. Gibbons, Christopher~D. Hamley, Chung-Yu Shih, and Michael~S.
  Chapman.
\newblock Nondestructive fluorescent state detection of single neutral atom
  qubits.
\newblock {\em Phys. Rev. Lett.}, 106:133002, 2011.

\bibitem{Ise10}
L.~Isenhower, E.~Urban, X.~L. Zhang, A.~T. Gill, T.~Henage, T.~A. Johnson,
  T.~G. Walker, and M.~Saffman.
\newblock Demonstration of a neutral atom controlled-{NOT} quantum gate.
\newblock {\em Phys. Rev. Lett.}, 104:010503, 2010.

\bibitem{Olm10}
S.~Olmschenk, R.~Chicireanu, K.~D. Nelson, and J.~V. Porto.
\newblock Randomized benchmarking of atomic qubits in an optical lattice.
\newblock {\em New J. Phys.}, 12:113007, 2010.

\bibitem{She10}
J.~F. Sherson, C.~Weitenberg, M.~Endres, M.~Cheneau, I.~Bloch, and S.~Kuhr.
\newblock Single-atom-resolved fluorescence imaging of an atomic mott
  insualtor.
\newblock {\em Nature}, 467:68, 2010.

\bibitem{Wei10}
H.~Weimer, M.~Muller, Zoller~P. Lesanovsky, I., and H.P. Buchler.
\newblock A {R}ydberg quantum simulator.
\newblock {\em Nature Physics}, 6:382, 2010.

\bibitem{Wil10}
T.~Wilk, A.~Ga\"etan, C.~Evellin, J.~Wolters, Y.~Miroshnychenko, P.~Grangier,
  and A.~Browaeys.
\newblock Entanglement of two individual neutral atoms using {R}ydberg
  blockade.
\newblock {\em Phys. Rev. Lett.}, 104:010502, 2010.

\bibitem{Zha10}
X.~L. Zhang, L.~Isenhower, A.~T. Gill, T.~G. Walker, and M.~Saffman.
\newblock Deterministic entanglement of two neutral atoms via {R}ydberg
  blockade.
\newblock {\em Phys. Rev. A}, 82:030306, 2010.

\bibitem{Kay06}
A.~Kay, J.~K. Pachos, and C.~S. Adams.
\newblock Graph-state preparation and quantum computation with global
  addressing of optical lattices.
\newblock {\em Phys. Rev. A}, 73:022310, 2006.

\bibitem{Tur98}
Q.~A. Turchette, C.~S. Wood, B.~E. King, C.~J. Myatt, D.~Leibfried, W.~M.
  Itano, C.~Monroe, and D.~J. Wineland.
\newblock Deterministic entanglement of two trapped ions.
\newblock {\em Phys. Rev. Lett.}, 81:3631, 1998.

\bibitem{Cir00}
J.~I. Cirac and P.~Zoller.
\newblock A scalable quantum computer with ions in an array of microtraps.
\newblock {\em Nature}, 404:579--581, 2000.

\bibitem{Sac00}
C.~A. Sackett, D.~Kielpinski, B.~E. King, C.~Langer, V.~Meyer, C.~J. Myatt,
  M.~Rowe, Q.~A. Turchette, W.~M. Itano, D.~J. Wineland, and C.~Monroe.
\newblock Experimental entanglement of four particles.
\newblock {\em Nature}, 404:256--259, 2000.

\bibitem{Sch03}
F.~Schmidt-Kaler, H.~Häffner, M.~Riebe, S.~Gulde, G.~P.~T. Lancaster,
  T.~Deuschle, C.~Becher, C.~F. Roos, J.~Eschner, and R.~Blatt.
\newblock Realization of the {Cirac--Zoller controlled-NOT} quantum gate.
\newblock {\em Nature}, 422:408, 2003.

\bibitem{Gul03}
S.~Gulde, M.~Riebe, G.~P.~T. Lancaster, C.~Becher, J.~Eschner, H.~H\"affner,
  I.~L. Chuang, R.~Blatt, and F.~Schmidt-Kaler.
\newblock Implementation of the {Deutsch–-Jozsa} algorithm on an ion-trap
  quantum computer.
\newblock {\em Nature}, 421:48--50, 2003.

\bibitem{Lei03}
D.~Leibfried, B.~DeMarco, V.~Meyer, D.~Lucas, M.~Barrett, J.~Britton, W.~M.
  Itano, B.~Jelenkovic, C.~Langer, T.~Rosenband, and D.~J. Wineland.
\newblock Experimental demonstration of a robust, high-fidelity geometric two
  ion-qubit phase gate.
\newblock {\em Nature}, 422:412, 2003.

\bibitem{Bar04}
M.~D. Barrett, J.~Chiaverini, T.~Schaetz, J.~Britton, W.~M. Itano, J.~D. Jost,
  E.~Knill, C.~Langer, D.~Leibfried, R.~Ozeri, and D.~J. Wineland.
\newblock Deterministic quantum teleportation of atomic qubits.
\newblock {\em Nature}, 429:737--739, 2004.

\bibitem{Chi04}
J.~Chiaverini, D.~Leibfried, T.~Schaetz, M.~D. Barrett, R.~B. Blakestad,
  J.~Britton, W.~M. Itano, J.~D. Jost, E.~Knill, C.~Langer, R.~Ozeri, and D.~J.
  Wineland.
\newblock Realization of quantum error correction.
\newblock {\em Nature}, 432:602--605, 2004.

\bibitem{Rie04}
M.~Riebe, H.~H\"affner, C.~F. Roos, W.~Hansel, J.~Benhelm, G.~P.~T. Lancaster,
  T.~W. Korber, C.~Becher, F.~Schmidt-Kaler, D.~F.~V. James, and R.~Blatt.
\newblock Deterministic quantum teleportation with atoms.
\newblock {\em Nature}, 429:734--737, 2004.

\bibitem{Chi05}
J.~Chiaverini, J.~Britton, D.~Leibfried, E.~Knill, M.~D. Barrett, R.~B.
  Blakestad, W.~M. Itano, J.~D. Jost, C.~Langer, R.~Ozeri, T.~Schaetz, and
  D.~J. Wineland.
\newblock Implementation of the semiclassical quantum fourier transform in a
  scalable system.
\newblock {\em Science}, 308:997--1002, 2005.

\bibitem{Hae05}
H.~H\"affner, W.~H\"ansel, C.~F. Roos, J.~Benhelm, D.~Chekalkar, M.~Chwalla,
  T.~K\"orber, U.~D. Rapol, M.~Riebe, P.~O. Schmidt, C.~Becher, O.~G\"uhne,
  W.~D\"ur, and R.~Blatt.
\newblock Scalable multiparticle entanglement of trapped ions.
\newblock {\em Nature}, 438:643--646, 2005.

\bibitem{Haf05}
H.~H\"affner, F.~Schmidt-Kaler, W.~Haensel, C.~F. Roos, T.~Koerber, M.~Chwalla,
  M.~Riebe, J.~Benhelm, U.D. Rapol, C.~Becher, and R.~Blatt.
\newblock Robust entanglement.
\newblock {\em Appl. Phys. B}, 81:151, 2005.

\bibitem{Lei05}
D.~Leibfried, E.~Knill, S.~Seidelin, J.~Britton, R.~B. Blakestad,
  J.~Chiaverini, D.~B. Hume, W.~M. Itano, J.~D. Jost, C.~Langer, R.~Ozeri,
  R.~Reichle, and D.~J. Wineland.
\newblock Creation of a six-atom {S}chr\"odinger cat state.
\newblock {\em Nature}, 438:639--642, 2005.

\bibitem{Rie06}
M.~Riebe, K.~Kim, P.~Schindler, T.~Monz, P.~O. Schmidt, T.~K. Korber,
  W.~Hansel, H.~H\"affner, C.~F. Roos, and R.~Blatt.
\newblock Process tomography of ion trap quantum gates.
\newblock {\em Phys. Rev. Lett.}, 97:220407, 2006.

\bibitem{Sti06}
D.~Stick, W.~K. Hensinger, S.~Olmschenk, M.~J. Madsen, K.~Schwab, and
  C.~Monroe.
\newblock Ion trap in a semiconductor chip.
\newblock {\em Nature Physics}, 2:36--39, 2006.

\bibitem{Moe07}
D.~L. Moehring, P.~Maunz, S.~Olmschenk, K.~C. Younge, D.~N. Matsukevich, L.-M.
  Duan, and C.~Monroe.
\newblock Entanglement of single--atom quantum bits at a distance.
\newblock {\em Nature}, 449:68--71, 2007.

\bibitem{Ben08}
J.~Benhelm, G.~Kirchmair, C.~F. Roos, and R.~Blatt.
\newblock Towards fault--tolerant quantum computing with trapped ions.
\newblock {\em Nature Physics}, 4:463--466, 2008.

\bibitem{Kie08}
D.~Kielpinski.
\newblock Ion--trap quantum information processing: experimental status.
\newblock {\em Frontiers of Physics in China}, 3:365--381, 2008.

\bibitem{Han09}
D.~Hanneke, J.~P. Home, J.~D. Jost, J.~M. Amini, D.~Leibfried, and D.~J.
  Wineland.
\newblock Realization of a programmable two-qubit quantum processor.
\newblock {\em Nature Physics}, 6:13--16, 2010.

\bibitem{Myr08}
A.~H. Myerson, D.~J. Szwer, S.~C. Webster, D.~T.~C. Allcock, M.~J. Curtis,
  G.~Imreh, J.~A. Sherman, D.~N. Stacey, A.~M. Steane, and D.~M. Lucas.
\newblock High-fidelity readout of trapped-ion qubits.
\newblock {\em Phys. Rev. Lett.}, 100:200502, 2008.

\bibitem{Mon09}
T.~Monz, K.~Kim, W.~Hansel, M.~Riebe, A.~S. Villar, P.~Schindler, M.~Chwalla,
  M.~Hennrich, and R.~Blatt.
\newblock Realization of the quantum {T}offoli gate with trapped ions.
\newblock {\em Phys. Rev. Lett.}, 102:040501, 2009.

\bibitem{Moz09}
T.~Monz, K.~Kim, A.~S. Villar, P.~Schindler, M.~Chwalla, M.~Riebe, C.~F. Roos,
  H.~Haeffner, W.~Haensel, M.~Hennrich, and R.~Blatt.
\newblock Realization of universal ion trap quantum computation with
  decoherence free qubits.
\newblock {\em Phys. Rev. Lett.}, 103:200503, 2009.

\bibitem{Bur10}
A.~H. Burrell, D.~J. Szwer, S.~C. Webster, and D.~M. Lucas.
\newblock Scalable simultaneous multi-qubit readout with 99.99$\%$ single-shot
  fidelity.
\newblock {\em Phys. Rev. A}, 81:04030, 2010.

\bibitem{Cam10}
W.~C. Campbell, J.~Mizrahi, Q.~Quraishi, C.~Senko, D.~Hayes, D.~Hucul, D.~N.
  Matsukevich, P.~Maunz, and C.~Monroe.
\newblock Ultrafast gates for single atomic qubits.
\newblock {\em Phys. Rev. Lett.}, 105(9):090502, 2010.

\bibitem{Mon10}
T.~Monz, P.~Schindler, J.~T. Barreiro, M.~Chwalla, D.~Nigg, W.~A. Coish,
  M.~Harlander, W.~Haensel, M.~Hennrich, and R.~Blatt.
\newblock 14-qubit entanglement: Creation and coherence.
\newblock {\em Phys. Rev. Lett.}, 106:130506, 2011.

\bibitem{Nak99}
Y.~Nakamura, Y.~A. Pashkin, and J.~S. Tsai.
\newblock Coherent control of macroscopic quantum states in a
  single-{C}ooper-pair box.
\newblock {\em Nature}, 398:786, 1999.

\bibitem{Nak02}
Y.~Nakamura, Yu.~A. Pashkin, T.~Yamamoto, and J.~S. Tsai.
\newblock Charge echo in a {C}ooper-pair box.
\newblock {\em Phys. Rev. Lett.}, 88:047901, 2002.

\bibitem{Mar02}
J.~M. Martinis, S.~Nam, J.~Aumentado, and C.~Urbina.
\newblock Rabi oscillations in a large {J}osephson-junction qubit.
\newblock {\em Phys. Rev. Lett.}, 89:117901, 2002.

\bibitem{Vio02}
D.~Vion, A.~Aassime, A.~Cottet, P.~Joyez, H.~Pothier, C.~Urbina, D.~Esteve, and
  M.~H. Devoret.
\newblock Manipulating the quantum state of an electrical circuit.
\newblock {\em Science}, 296:886, 2002.

\bibitem{Chi03}
I.~Chiorescu, Y.~Nakamura, C.~J. P.~M. Harmans, and J.~E. Mooij.
\newblock Coherent quantum dynamics of a superconducting flux qubit.
\newblock {\em Science}, 299:1869, 2003.

\bibitem{Ber03}
A.~J. Berkley, H.~Xu, R.~C. Ramos, M.~A. Gubrud, F.~W. Strauch, P.~R. Johnson,
  J.~Anderson, A.~J. Dragt, C.~J. Lobb, and F.~C. Wellstood.
\newblock Entangled macroscopic quantum states in two superconducting qubits.
\newblock {\em Science}, 300:1548, 2003.

\bibitem{Yam03}
T.~Yamamoto, Y.~A. Pashkin, O.~Astafiev, Y.~Nakamura, and J.~S. Tsai.
\newblock Demonstration of conditional gate operation using superconducting
  charge qubits.
\newblock {\em Nature}, 425:941, 2003.

\bibitem{Wal04}
A.~Wallraff, D.~I. Schuster, A.~Blais, L.~Frunzio, R.~S. Huang, J.~Majer,
  S.~Kumar, S.~M. Girvin, and R.~J. Schoelkopf.
\newblock Strong coupling of a single photon to a superconducting qubit using
  circuit quantum electrodynamics.
\newblock {\em Nature}, 431:162, 2004.

\bibitem{McD05}
R.~McDermott, R.~W. Simmonds, Matthias Steffen, K.~B. Cooper, K.~Cicak, K.~D.
  Osborn, Seongshik Oh, D.~P. Pappas, and J.~M. Martinis.
\newblock Simultaneous state measurement of coupled {J}osephson phase qubits.
\newblock {\em Science}, 307:1299--1302, 2005.

\bibitem{Him06}
T.~Hime, P.~A. Reichardt, B.~L.~T. Plourde, T.~L. Robertson, C.-E. Wu, A.~V.
  Ustinov, and John Clarke.
\newblock Solid-state qubits with current-controlled coupling.
\newblock {\em Science}, 314:1427--1429, 2006.

\bibitem{Gra06}
M.~Grajcar, A.~Izmalkov, S.~H.~W. van~der Ploeg, S.~Linzen, T.~Plecenik, Th.
  Wagner, U.~Hubner, E.~Il'ichev, H.-G. Meyer, A.~Yu. Smirnov, Peter~J. Love,
  Alec~Maassen van~den Brink, M.~H.~S. Amin, S.~Uchaikin, and A.~M. Zagoskin.
\newblock Four-qubit device with mixed couplings.
\newblock {\em Physical Review Letters}, 96:047006, 2006.

\bibitem{Ste06}
M.~Steffen, M.~Ansmann, R.~C. Bialczak, N.~Katz, E.~Lucero, R.~McDermott,
  M.~Neeley, E.~M. Weig, A.~N. Cleland, and J.~M. Martinis.
\newblock Measurement of the entanglement of two superconducting qubits via
  state tomography.
\newblock {\em Science}, 313:1423--1425, 2006.

\bibitem{Yos06}
F.~Yoshihara, K.~Harrabi, A.~O. Niskanen, Y.~Nakamura, and J.~S. Tsai.
\newblock Decoherence of flux qubits due to $1/f$ flux noise.
\newblock {\em Phys. Rev. Lett.}, 97:167001, 2006.

\bibitem{Har07}
R.~Harris, A.~J. Berkley, M.~W. Johnson, P.~Bunyk, S.~Govorkov, M.~C. Thom,
  S.~Uchaikin, A.~B. Wilson, J.~Chung, E.~Holtham, J.~D. Biamonte, A.~Yu.
  Smirnov, M.~H.~S. Amin, and A.~Maassen van~den Brink.
\newblock Sign- and magnitude-tunable coupler for superconducting flux qubits.
\newblock {\em Phys. Rev. Lett.}, 98:177001, 2007.

\bibitem{Lup07}
A.~Lupascu, S.~Saito, T.~Picot, C.~J. P. M.~Harmans P.~C.~de Groot~and, and
  J.~E. Mooij.
\newblock Quantum non-demolition measurement of a superconducting two-level
  system.
\newblock {\em Nature Physics}, 3:119--125, 2007.

\bibitem{Maj07}
J.~Majer, J.~M. Chow, J.~M. Gambetta, Jens Koch, B.~R. Johnson, J.~A. Schreier,
  L.~Frunzio, D.~I. Schuster, A.~A. Houck, A.~Wallraff, A.~Blais, M.~H.
  Devoret, S.~M. Girvin, and R.~J. Schoelkopf.
\newblock Coupling superconducting qubits via a cavity bus.
\newblock {\em Nature}, 449:443--447, 2007.

\bibitem{Nis07}
A.~O. Niskanen, K.~Harrabi, F.~Yoshihara, Y.~Nakamura, S.~Lloyd, and J.~S.
  Tsai.
\newblock Quantum coherent tunable coupling of superconducting qubits.
\newblock {\em Science}, 316:723--726, 2007.

\bibitem{Pla07}
J.~H. Plantenberg, P.~C. de~Groot, C.~J. P.~M. Harmans, and J.~E. Mooij.
\newblock Demonstration of controlled-{NOT} quantum gates on a pair of
  superconducting quantum bits.
\newblock {\em Nature}, 447:836, 2007.

\bibitem{Sil07}
M.~A. Sillanp\"a\"a, J.~I. Park, and R.~W. Simmonds.
\newblock Coherent quantum state storage and transfer between two phase qubits
  via a resonant cavity.
\newblock {\em Nature}, 449:438--442, 2007.

\bibitem{Pic08}
T.~Picot, A.~Lupascu, S.~Saito, C.~J. P.~M. Harmans, and J.~E. Mooij.
\newblock Role of relaxation in the quantum measurement of a superconducting
  qubit using a nonlinear oscillator.
\newblock {\em Phys. Rev. B}, 78:132508, 2008.

\bibitem{Sch08}
J.~A. Schreier, A.~A. Houck, Jens Koch, D.~I. Schuster, B.~R. Johnson, J.~M.
  Chow, J.~M. Gambetta, J.~Majer, L.~Frunzio, M.~H. Devoret, S.~M. Girvin, and
  R.~J. Schoelkopf.
\newblock Suppressing charge noise decoherence in superconducting charge
  qubits.
\newblock {\em Phys. Rev. B}, 77:180502, 2008.

\bibitem{Sho08}
R.~J. Schoelkopf and S.~M. Girvin.
\newblock Wiring up quantum systems.
\newblock {\em Nature}, 451:664--669, 2008.

\bibitem{Ans09}
M.~Ansmann, H.~Wang, R.~C. Bialczak, M.~Hofheinz, E.~Lucero, M.~Neeley, A.~D.
  O'Connell, D.~Sank, M.~Weides, J.~Wenner, A.~N. Cleland, and J.~M. Martinis.
\newblock Violation of {B}ell's inequality in josephson phase qubits.
\newblock {\em Nature}, 461:504, 2009.

\bibitem{DiC09}
L.~DiCarlo, J.~M. Chow, J.~M. Gambetta, L.~S. Bishop, B.~R. Johnson, D.~I.
  Schuster, J.~Majer, A.~Blais, L.~Frunzio, S.~M. Girvin, and R.~J. Schoelkopf.
\newblock Demonstration of two--qubit algorithms with a superconducting quantum
  processor.
\newblock {\em Nature}, 460:240--244, 2009.

\bibitem{Man09}
V.~E. Manucharyan, J.~Koch, L.~I. Glazman, and M.~H. Devoret.
\newblock Fluxonium: Single {C}ooper-pair circuit free of charge offsets.
\newblock {\em Science}, 326:113, 2009.

\bibitem{Nee10}
M.~Neeley, R.~C. Bialczak, M.~Lenander, E.~Lucero, M.~Mariantoni, A.~D.
  O'Connell, D.~Sank, H.~Wang, M.~Weides, J.~Wenner, Y.~Yin, T.~Yamamoto, A.~N.
  Cleland, and J.~M. Martinis.
\newblock Generation of three-qubit entangled states using superconducting
  phase qubits.
\newblock {\em Nature}, 467:570, 2010.

\bibitem{diC10}
L.~DiCarlo, M.~D. Reed, L.~Sun, B.~R. Johnson, J.~M. Chow, J.~M. Gambetta,
  L.~Frunzio, S.~M. Girvin, M.~H. Devoret, and R.~J. Schoelkopf.
\newblock Preparation and measurement of three-qubit entanglement in a
  superconducting circuit.
\newblock {\em Nature}, 467:574, 2010.

\bibitem{Sun10}
G.~Sun, X.~Wen, B.~Mao, J.~Chen, Y.~Yu, P.~Wu, and S.~Han.
\newblock Tunable quantum beam splitters for coherent manipulation of a
  solid-state tripartite qubit system.
\newblock {\em Nature Communications}, 1:51, 2010.

\bibitem{Har10}
R.~Harris, M.~W. Johnson, T.~Lanting, A.~J. Berkley, J.~Johansson, P.~Bunyk,
  E.~Tolkacheva, E.~Ladizinsky, N.~Ladizinsky, T.~Oh, F.~Cioata, I.~Perminov,
  P.~Spear, C.~Enderud, C.~Rich, S.~Uchaikin, M.~C. Thom, E.~M. Chapple,
  J.~Wang, B.~Wilson, M.~H.~S. Amin, N.~Dickson, K.~Karimi, B.~Macready,
  C.~J.~S. Truncik, and G.~Rose.
\newblock Experimental investigation of an eight-qubit unit cell in a
  superconducting optimization processor.
\newblock {\em Phys. Rev. B}, 82(2):024511, 2010.

\bibitem{OCo10}
A.~D. O'Connell, M.~Hofheinz, M.~Ansmann, R.~C. Bialczak, M.~Lenander,
  E.~Lucero, M.~Neeley, D.~Sank, H.~Wang, M.~Weides, J.~Wenner, J.~M. Martinis,
  and A.~N. Cleland.
\newblock Quantum ground state and single-phonon control of a mechanical
  resonator.
\newblock {\em Nature}, 464:697, 2010.

\bibitem{Pal10}
A.~Palacios-Laloy, F.~Mallet, F.~Nguyen, P.~Bertet, D.~Vion, D.~Esteve, and
  A.~N. Korotkov.
\newblock Experimental violation of a {B}ell's inequality in time with weak
  measurement.
\newblock {\em Nature Physics}, 7:442–447, 2010.

\bibitem{Ste10}
M.~Steffen, S.~Kumar, D.P. DiVincenzo, J.R. Rozen, G.~A. Keefe, M.~B. Rothwell,
  and M.~B. Ketchen.
\newblock High coherence hybrid superconducting qubit.
\newblock {\em Phys. Rev. Lett.}, 105:100502, 2010.

\bibitem{Mar11}
M.~Mariantoni.
\newblock private communication, 2011.

\bibitem{Byl11}
J.~Bylander, S.~Gustavsson, F.~Yan, F.~Yoshihara, K.~Harrabi, G.~Fitch, D.~G.
  Cory, Y.~Nakamura, J.-S. Tsai, and W.~D. Oliver.
\newblock Noise spectroscopy through dynamical decoupling with a
  superconducting flux qubit.
\newblock {\em Nature Physics}, 7:565–570, 2011.

\bibitem{Martinis}
M.~Neeley, M.~Ansmann, R.~C. Bialczak, M.~Hofheinz, E.~Lucero, A.~D. O'Connell,
  D.~Sank, H.~Wang, J.~Wenner, A.~N. Cleland, M.~R. Geller, and J.~M. Martinis.
\newblock Emulation of a quantum spin with a superconducting phase qudit.
\newblock {\em Science}, 325:722 -- 725, 2009.

\bibitem{Tyr11}
A.~M. Tyryshkin, S.~Tojo, J.~J.~L. Morton, H.~Riemann, N.~V. Abrosimov,
  P.~Becker, H.-J. Pohl, T.~Schenkel, M.~L.~W. Thewalt, K.~M. Itoh, and S.~A.
  Lyon.
\newblock Electron spin coherence exceeding seconds in high purity silicon.
\newblock 2011.

\bibitem{Mak01}
Y.~Makhlin, G.~Schoen, and A.~Shnirman.
\newblock Quantum-state engineering with {J}osephson-junction devices.
\newblock {\em Rev. Mod. Phys.}, 73:357, 2001.

\bibitem{You11}
J.~Q. You and F.~Nori.
\newblock Atomic physics and quantum optics using superconducting circuits.
\newblock {\em Nature}, 474:589, 2011.

\bibitem{Hof09}
M.~Hofheinz, H.~Wang, M.~Ansmann, R.~C. Bialczak, E.~Lucero, M.~Neeley, A.~D.
  O'Connell, D.~Sank, J.~Wenner, John~M. Martinis, and A.~N. Cleland.
\newblock Synthesizing arbitrary quantum states in a superconducting resonator.
\newblock {\em Nature}, 459:546--549, 2009.

\bibitem{Nor09}
F.~Nori.
\newblock Quantum football.
\newblock {\em Science}, 325:689, 2009.

\bibitem{Los98}
D.~Loss and D.~P. DiVincenzo.
\newblock Quantum computation with quantum dots.
\newblock {\em Phys. Rev. A}, 57:120--126, 1998.

\bibitem{Tyr03}
A.~M. Tyryshkin, S.~A. Lyon, A.~V. Astashkin, and A.~M. Raitsimring.
\newblock Electron spin relaxation times of phosphorus donors in silicon.
\newblock {\em Phys. Rev. B}, 68:193207, 2003.

\bibitem{Jel04}
F.~Jelezko, T.~Gaebel, I.~Popa, A.~Gruber, and J.~Wrachtrup.
\newblock Observation of coherent oscillations in a single electron spin.
\newblock {\em Phys. Rev. Lett.}, 92:076401, 2004.

\bibitem{Jele04}
F.~Jelezko, T.~Gaebel, I.~Popa, M.~Domhan, A.~Gruber, and J.~Wrachtrup.
\newblock Observation of coherent oscillation of a single nuclear spin and
  realization of a two-qubit conditional quantum gate.
\newblock {\em Phys. Rev. Lett.}, 93:130501, 2004.

\bibitem{Pet05}
J.~R. Petta, A.~C. Johnson, J.~M. Taylor, E.~A. Laird, A.~Yacoby, M.~D. Lukin,
  C.~M. Marcus, M.~P. Hanson, and A.~C. Gossard.
\newblock Coherent manipulation of coupled electron spins in semiconductor
  quantum dots.
\newblock {\em Science}, 309:2180--2184, 2005.

\bibitem{Chi06}
L.~Childress, M.~V. Gurudev~Dutt, J.~M. Taylor, A.~S. Zibrov, F.~Jelezko,
  J.~Wrachtrup, P.~R. Hemmer, and M.~D. Lukin.
\newblock Coherent dynamics of coupled electron and nuclear spin qubits in
  diamond.
\newblock {\em Science}, 314(5797):281--285, 2006.

\bibitem{Han06}
R.~Hanson, F.~M. Mendoza, R.~J. Epstein, and D.~D. Awschalom.
\newblock Polarization and readout of coupled single spins in diamond.
\newblock {\em Phys. Rev. Lett.}, 97:087601, 2006.

\bibitem{Kop06}
F.~H.~L. Koppens, C.~Buizert, K.~J. Tielrooij, I.~T. Vink, K.~C. Nowack,
  T.~Meunier, L.~P. Kouwenhoven, and L.~M.~K. Vandersypen.
\newblock Driven coherent oscillations of a single electron spin in a quantum
  dot.
\newblock {\em Nature}, 442:766--771, 2006.

\bibitem{Steg06}
A.~R. Stegner, C.~Boehme, H.~Huebl, M.~Stutzmann, K.~Lips, and M.~S. Brandt.
\newblock Electrical detection of coherent 31p spin quantum states.
\newblock {\em Nature Physics}, 2:835 -- 838, 2006.

\bibitem{Dut07}
M.~V. Gurudev~Dutt, L.~Childress, L.~Jiang, E.~Togan, J.~Maze, F.~Jelezko,
  A.~S. Zibrov, P.~R. Hemmer, and M.~D. Lukin.
\newblock Quantum register based on individual electronic and nuclear spin
  qubits in diamond.
\newblock {\em Science}, 316:1312--1316, 2007.

\bibitem{Han07}
R.~Hanson, L.~P. Kouwenhoven, J.~R. Petta, S.~Tarucha, and L.~M.~K.
  Vandersypen.
\newblock Spins in few-electron quantum dots.
\newblock {\em Reviews of Modern Physics}, 79:1217, 2007.

\bibitem{Mik07}
M.~H. Mikkelsen, J.~Berezovsky, L.~A.~Coldren N.~G. Stoltz~and, and D.~D.
  Awschalom.
\newblock Optically detected coherent spin dynamics of a single electron in a
  quantum dot.
\newblock {\em Nature Physics}, 3:770--773, 2007.

\bibitem{Now07}
K.~C. Nowack, F.~H.~L. Koppens, Y.~V. Nazarov, and L.~M.~K. Vandersypen.
\newblock Coherent control of a single electron spin with electric fields.
\newblock {\em Science}, 318:1430--1433, 2007.

\bibitem{Xu07}
X.~Xu, B.~Sun, P.~R. Berman, D.~G. Steel, A.~S. Bracker, D.~Gammon, and L.~J.
  Sham.
\newblock Coherent optical spectroscopy of a strongly driven quantum dot.
\newblock {\em Science}, 317:929--932, 2007.

\bibitem{Ama08}
S.~Amasha, K.~MacLean, I.~P. Radu, D.~M. Zumbuhl, M.~A. Kastner, M.~P. Hanson,
  and A.~C. Gossard.
\newblock Electrical control of spin relaxation in a quantum dot.
\newblock {\em Phys. Rev. Lett.}, 100:046803, 2008.

\bibitem{Ber08}
J.~Berezovsky, M.~H. Mikkelsen, N.~G. Stoltz, L.~A. Coldren, and D.~D.
  Awschalom.
\newblock Picosecond coherent optical manipulation of a single electron spin in
  a quantum dot.
\newblock {\em Science}, 320:349--352, 2008.

\bibitem{Chi08}
L.~Chirolli and G.~Burkard.
\newblock Decoherence in solid state qubits.
\newblock {\em Advances in Physics}, 57:225, 2008.

\bibitem{Ger08}
B.~D. Gerardot, D.~Brunner, P.~A. Dalgarno, P.~Ohberg, S.~Seidl, M.~Kroner,
  K.~Karrai, N.~G. Stoltz, P.~M. Petroff, and R.~J. Warburton.
\newblock Optical pumping of a single hole spin in a quantum dot.
\newblock {\em Nature}, 451:441--444, 2008.

\bibitem{Neu08}
P.~Neumann, N.~Mizuochi, F.~Rempp, P.~Hemmer, H.~Watanabe, S.~Yamasaki,
  V.~Jacques, T.~Gaebel, F.~Jelezko, and J.~Wrachtrup.
\newblock Multipartite entanglement among single spins in diamond.
\newblock {\em Science}, 320:1326--1329, 2008.

\bibitem{Bal09}
G.~Balasubramanian, P.~Neumann, D.~Twitchen, M.~Markham, R.~Kolesov,
  N.~Mizuochi, J.~Isoya, J.~Achard, J.~Beck, J.~Tissler, V.~Jacques, P.R.
  Hemmer, F.~Jelezko, and J.~Wrachtrup.
\newblock Ultralong spin coherence time in isotopically engineered diamond.
\newblock {\em Nature Materials}, 8:383--387, 2009.

\bibitem{Bar09}
C.~Barthel, D.~J. Reilly, C.~M. Marcus, M.~P. Hanson, and A.~C. Gossard.
\newblock Rapid single-shot measurement of a singlet-triplet qubit.
\newblock {\em Phys. Rev. Lett.}, 103:160503, 2009.

\bibitem{Fol09}
S.~Foletti, H.~Bluhm, D.~Mahalu, V.~Umansky, and A.~Yacoby.
\newblock Universal quantum control of two-electron spin quantum bits using
  dynamic nuclear polarization.
\newblock {\em Nature Physics}, 5:903, 2009.

\bibitem{Twa09}
J.~Twamley and S.~D. Barrett.
\newblock A superconducting cavity bus for single {N}itrogen {V}acancy defect
  centres in diamond.
\newblock {\em Phys. Rev. B}, 81:241202, 2010.

\bibitem{Bar10}
C.~Barthel, J.~Medford, C.~M. Marcus, M.~P. Hanson, and A.~C. Gossard.
\newblock Interlaced dynamical decoupling and coherent operation of a
  singlet-triplet qubit.
\newblock {\em Phys. Rev. Lett.}, 105(26):266808, 2010.

\bibitem{Blu10}
H.~Bluhm, S.~Foletti, D.~Mahalu, V.~Umansky, and A.~Yacoby.
\newblock Enhancing the coherence of a spin qubit by operating it as a feedback
  loop that controls its nuclear spin bath.
\newblock {\em Phys. Rev. Lett.}, 105:216803, 2010.

\bibitem{Lan10}
G.~de~Lange, Z.~H. Wang, D.~Ristè, V.~V. Dobrovitski, and R.~Hanson.
\newblock Universal dynamical decoupling of a single solid-state spin from a
  spin bath.
\newblock {\em Science}, 330:60, 2010.

\bibitem{Nad10}
S.~Nadj-Perge, S.~M. Frolov, E.~P. A.~M. Bakkers, and L.~P. Kouwenhoven.
\newblock Spin--orbit qubit in a semiconductor nanowire.
\newblock {\em Nature}, 468:1084, 2010.

\bibitem{Blu11}
H.~Bluhm, S.~Foletti, I.~Neder, M.~Rudner, D.~Mahalu, V.~Umansky, and
  A.~Yacoby.
\newblock Dephasing time of {G}a{A}s electron-spin qubits coupled to a nuclear
  bath exceeding 200$\mu$s.
\newblock {\em Nature Physics}, 7:109, 2011.

\bibitem{Mor10}
A.~Morello, J.~J. Pla, F.~A. Zwanenburg, K.~W. Chan, K.~Y. Tan, H.~Huebl,
  M.~Mottonen, C.~D. Nugroho, C.~Yang, J.~A. van Donkelaar, A.~D.~C. Alves,
  D.~N. Jamieson, C.~C. Escott, L.~C.~L. Hollenberg, R.~G. Clark, and A.~S.
  Dzurak.
\newblock Single-shot readout of an electron spin in silicon.
\newblock {\em Nature}, 467:687, 2010.

\bibitem{MCC10}
D.~R. McCamey, J.~Van~Tol, G.~W. Morley, and C.~Boehme.
\newblock Electronic spin storage in an electrically readable nuclear spin
  memory with a lifetime $ >$100 seconds.
\newblock {\em Science}, 330:1652--1656, 2010.

\bibitem{Wit10}
W.~M. Witzel, M.~S. Carroll, A.~Morello, L.~Cywi\ifmmode~\acute{n}\else
  \'{n}\fi{}ski, and S.~Das~Sarma.
\newblock Electron spin decoherence in isotope-enriched silicon.
\newblock {\em Phys. Rev. Lett.}, 105:187602, 2010.

\bibitem{Sims10}
C.~B. Simmons, J.~R. Prance, B.~J. Van~Bael, Teck~Seng Koh, Zhan Shi, D.~E.
  Savage, M.~G. Lagally, R.~Joynt, Mark Friesen, S.~N. Coppersmith, and M.~A.
  Eriksson.
\newblock Tunable spin loading and $t1$ of a silicon spin qubit measured by
  single-shot readout.
\newblock {\em Phys. Rev. Lett.}, 106:156804, 2011.

\bibitem{Sim11}
S.~Simmons, R.~M. Brown, H.~Riemann, N.~V. Abrosimov, P.~Becker, H.~J. Pohl,
  L.~W. Thewalt, and J.~J.~L. Morton.
\newblock Entanglement in a solid-state spin ensemble.
\newblock {\em Nature}, 470:69–72, 2011.

\bibitem{Sut08}
D.~Suter and T.~S. Mahesh.
\newblock Spins as qubits: Quantum information processing by nuclear magnetic
  resonance.
\newblock {\em J. Chem. Phys.}, 128:052206, 2008.

\bibitem{Pen05}
X.~Peng, Z.~Liao, N.~Xu, G.~Qin, X.~Zhou, D.~Suter, and J.~Du.
\newblock Quantum adiabatic algorithm for factorization and its experimental
  implementation.
\newblock {\em Phys. Rev. Lett.}, 101:220405, 2008.

\bibitem{Neg06}
C.~Negrevergne, T.~S. Mahesh, C.~A. Ryan, M.~Ditty, F.~Cyr-Racine, W.~Power,
  N.~Boulant, T.~Havel, D.~G. Cory, and R.~Laflamme.
\newblock Benchmarking quantum control methods on a 12-qubit system.
\newblock {\em Phys. Rev. Lett.}, 96:170501, 2006.

\bibitem{Van01}
L.~M.~K. Vandersypen, M.~Steffen, G.~Breyta, C.~S. Yannoni, M.~H. Sherwood, and
  I.~L. Chuang.
\newblock Experimental realization of {S}hor's quantum factoring algorithm
  using nuclear magnetic resonance.
\newblock {\em Nature}, 414:883--887, 2001.

\bibitem{Gra08}
M.~Grajcar, S.~H.~W. van~der Ploeg, A.~Izmalkov, E.~Il'ichev, H.-G. Meyer,
  A.~Fedorov, A.~Shnirman, and Gerd Schon.
\newblock Sisyphus cooling and amplification by a superconducting qubit.
\newblock {\em Nature Physics}, 4:612--616, 2008.

\bibitem{Nor08}
F.~Nori.
\newblock Atomic physics with a circuit.
\newblock {\em Nature Physics}, 4:589, 2008.

\bibitem{Lan05}
C.~Langer, R.~Ozeri, J.~D. Jost, J.~Chiaverini, B.~DeMarco, A.~Ben-Kish, R.~B.
  Blakestad, J.~Britton, D.~B. Hume, W.~M. Itano, D.~Leibfried, R.~Reichle,
  T.~Rosenband, T.~Schaetz, P.~O. Schmidt, and D.~J. Wineland.
\newblock Long-lived qubit memory using atomic ions.
\newblock {\em Phys. Rev. Lett.}, 95:060502, 2005.

\bibitem{Kim08}
H.~J. Kimble.
\newblock The quantum internet.
\newblock {\em Nature}, 453:1023--1030, 2008.

\bibitem{Col07}
Y.~Colombe, T.~Steinmetz, G.~Dubois, F.~Linke, D.~Hunger, and J.~Reichel.
\newblock Strong atom--field coupling for {B}ose–-{E}instein condensates in an
  optical cavity on a chip.
\newblock {\em Nature}, 450:272--276, 2007.

\bibitem{Her09}
P.~F. Herskind, A.~Dantan, J.~P. Marler, M.~Albert, and M.~Drewsen.
\newblock Realization of collective strong coupling with ion {C}oulomb crystals
  in an optical cavity.
\newblock {\em Nature Physics}, 5:494--498, 2009.

\bibitem{Eng07}
D.~Englund, A.~Faraon, I.~Fushman, N.~Stoltz, P.~Petroff, and J.~Vuckovic.
\newblock Controlling cavity reflectivity with a single quantum dot.
\newblock {\em Nature}, 450:857--861, 2007.

\bibitem{OB07}
J.~L. O'Brien.
\newblock Optical quantum computing.
\newblock {\em Science}, 318:1567--1570, 2007.

\bibitem{Tia04}
L.~Tian, P.~Rabl, R.~Blatt, and P.~Zoller.
\newblock Interfacing quantum-optical and solid-state qubits.
\newblock {\em Phys. Rev. Lett.}, 92:247902, 2004.

\bibitem{Tia05}
L.~Tian, R.~Blatt, and P.~Zoller.
\newblock Scalable ion trap quantum computing without moving ions.
\newblock {\em Eur. Phys. J. D}, 32:201--208, 2005.

\bibitem{Ver08}
J.~Verdu, H.~Zoubi, Ch. Koller, J.~Majer, H.~Ritsch, and J.~Schmiedmayer.
\newblock Strong magnetic coupling of an ultracold gas to a superconducting
  waveguide cavity.
\newblock {\em Phys. Rev. Lett.}, 103:043603, 2009.

\bibitem{Pet09}
D.~Petrosyan, G.~Bensky, G.~Kurizki, I.~Mazets, J.~Majer, and J.~Schmiedmayer.
\newblock Reversible state transfer between superconducting qubits and atomic
  ensembles.
\newblock {\em Phys. Rev. A}, 79:040304, 2009.

\bibitem{Zip10}
C.~Zipkes, S.~Palzer, C.~Sias, and M.~Kohl.
\newblock A trapped single ion inside a {Bose-Einstein} condensate.
\newblock {\em Nature}, 464:388, 2010.

\bibitem{Doe10}
H.~Doerk, Z.~Idziaszek, and T.~Calarco.
\newblock Atom-ion quantum gate.
\newblock {\em Phys. Rev. A}, 81:012708, 2010.

\bibitem{Kub10}
Y.~Kubo, F.~R. Ong, P.~Bertet, D.~Vion, V.~Jacques, D.~Zheng, A.~Dr\'eau, J.-F.
  Roch, A.~Auffeves, F.~Jelezko, J.~Wrachtrup, M.~F. Barthe, P.~Bergonzo, and
  D.~Esteve.
\newblock Strong coupling of a spin ensemble to a superconducting resonator.
\newblock {\em Phys. Rev. Lett.}, 105:140502, 2010.

\bibitem{Sch10}
D.~I. Schuster, A.~P. Sears, E.~Ginossar, L.~DiCarlo, L.~Frunzio, J.~J.~L.
  Morton, H.~Wu, G.~A.~D. Briggs, B.~B. Buckley, D.~D. Awschalom, and R.~J.
  Schoelkopf.
\newblock High-cooperativity coupling of electron-spin ensembles to
  superconducting cavities.
\newblock {\em Phys. Rev. Lett.}, 105:140501, 2010.

\bibitem{Wu10}
H.~Wu, R.~E. George, J.~H. Wesenberg, K.~M\o{}lmer, D.~I. Schuster, R.~J.
  Schoelkopf, K.~M. Itoh, A.~Ardavan, J.~J.~L. Morton, and G.~A.~D. Briggs.
\newblock Storage of multiple coherent microwave excitations in an electron
  spin ensemble.
\newblock {\em Phys. Rev. Lett.}, 105:140503, 2010.

\bibitem{Sim10}
C.~Simon, M.~Afzelius, J.~Appel, A.~Boyer de~la Giroday, S.~J. Dewhurst,
  N.~Gisin, C.~Y. Hu, F.~Jelezko, S.~Kroll, J.~H. Muller, J.~Nunn, E.~Polzik,
  J.~Rarity, H.~de~Riedmatten, W.~Rosenfeld, A.~J. Shields, N.~Skold, R.~M.
  Stevenson, R.~Thew, I.~Walmsley, M.~Weber, H.~Weinfurter, J.~Wrachtrup, and
  R.~J. Young.
\newblock Quantum memories. a review based on the european integrated project
  qubit applications (qap).
\newblock {\em Eur. Phys. J. D}, 58:1, 2010.

\bibitem{Wall09}
M.~Wallquist, K.~Hammerer, P.~Rabl, M.~Lukin, and P.~Zoller.
\newblock Hybrid quantum devices and quantum engineering.
\newblock {\em Physica Scripta}, T137:014001, 2009.

\bibitem{Div95}
D.~P. DiVincenzo.
\newblock Quantum computation.
\newblock {\em Science}, 270:255--261, 1995.

\bibitem{Bul09}
I.~M. Buluta and F.~Nori.
\newblock Quantum simulators.
\newblock {\em Science}, 326:108 -- 111, 2009.

\bibitem{Kim11}
Z.~Kim, B.~Suri, V.~Zaretskey, S.~Novikov, K.~D. Osborn, A.~Mizel, F.~C.
  Wellstood, and B.~S. Palmer.
\newblock Decoupling a cooper-pair box to enhance the lifetime to 0.2 ms.
\newblock {\em Phys. Rev. Lett.}, 106:120501, 2011.

\bibitem{Cir95}
J.~I. Cirac and P.~Zoller.
\newblock Quantum computations with cold trapped ions.
\newblock {\em Phys. Rev. Lett.}, 74:4091, 1995.

\bibitem{Mol99}
K.~M$\rm{\o}$lmer and A.~S$\rm{\o}$rensen.
\newblock Multiparticle entanglement of hot trapped ions.
\newblock {\em Phys. Rev. Lett.}, 82:1835, 1999.

\end{thebibliography}

\end{document}